\documentclass{article}

\usepackage{microtype}
\usepackage{graphicx}
\usepackage{booktabs} 
\usepackage{amsmath,amssymb}
\usepackage{hyperref}
\usepackage{bm}
\usepackage{subcaption}
\usepackage{pifont}

\newcommand{\ieno}{\textit{i}.\textit{e}.} 

\newcommand{\egno}{\textit{e}.\textit{g}.} 

\usepackage[accepted]{icml2021}

\icmltitlerunning{Soft then Hard: Rethinking the Quantization in Neural Image Compression}

\begin{document}

\twocolumn[
\icmltitle{Soft then Hard: Rethinking the Quantization in Neural Image Compression}



\icmlsetsymbol{equal}{*}

\begin{icmlauthorlist}
\icmlauthor{Zongyu Guo}{USTC}
\icmlauthor{Zhizheng Zhang}{USTC}
\icmlauthor{Runsen Feng}{USTC}
\icmlauthor{Zhibo Chen}{USTC}
\end{icmlauthorlist}

\icmlaffiliation{USTC}{University of Science and Technology of China}

\icmlcorrespondingauthor{Zongyu Guo}{guozy@mail.ustc.edu.cn}
\icmlcorrespondingauthor{Zhibo Chen}{chenzhibo@ustc.edu.cn}

\icmlkeywords{Machine Learning, ICML}

\vskip 0.3in
]


\printAffiliationsAndNotice{}  

\begin{abstract}

Quantization is one of the core components in lossy image compression. For neural image compression, end-to-end optimization requires differentiable approximations of quantization, which can generally be grouped into three categories: additive uniform noise, straight-through estimator and soft-to-hard annealing. Training with additive uniform noise approximates the quantization error variationally but suffers from the train-test mismatch. The other two methods do not encounter this mismatch but, as shown in this paper, hurt the rate-distortion performance since the latent representation ability is weakened. We thus propose a novel \emph{soft-then-hard} quantization strategy for neural image compression that first learns an expressive latent space softly, then closes the train-test mismatch with hard quantization. In addition, beyond the fixed integer quantization, we apply scaled additive uniform noise to adaptively control the quantization granularity by deriving a new variational upper bound on actual rate. Experiments demonstrate that our proposed methods are easy to adopt, stable to train, and highly effective especially on complex compression models.

\end{abstract}

\section{Introduction}
\label{section1}

Lossy image compression is a fundamental technique for image transmission and storage. Over the past several years, deep learning methods are reshaping this field. Despite the short history, learned image compression schemes \cite{toderici2017full,rippel2017real,balle2018variational,li2018learning,minnen2018joint,lee2019context,cheng2020learned} have surpassed almost all classical standards in terms of rate-distortion performance. Moreover, neural image compression is promising to be more perceptual friendly \cite{blau2019rethinking,mentzer2020high}.

Quantization is one of the key challenges for neural image compression.
Since the gradient of quantization is zero almost everywhere, it makes the standard back-propagation inapplicable. 
Although some recent works try to forgo the quantization process entirely \cite{havasi2019minimal,flamich2020compressing}, these methods are computationally costly and statistically inefficient \cite{agustsson2020universally}. Therefore, quantization remains indispensable for designing an efficient neural image codec. To enable end-to-end optimization, a popular approach is to train with additive uniform noise to approximate the test-time quantization \cite{balle2016end}.
However, this method introduces stochasticity during training, leading to the train-test mismatch and thus hurting the rate-distortion performance in this way.

Other competitive alternatives for quantization include straight-through estimator (STE) with its variants \cite{bengio2013estimating,theis2017lossy,mentzer2018conditional} and recently, soft-to-hard annealing \cite{agustsson2017soft,yang2020improving,agustsson2020universally}, both of which avoid the mismatch issue. 
In this paper, we introduce a new analysis of these three quantization methods and argue that:
\begin{itemize}
\item Training with STE or soft-to-hard annealing is equal to optimizing a deterministic autoencoder, in which it is hard to learn a smooth latent space due to the lack of regularization term at training \cite{ghosh2019variational}.
\item STE-based or annealing-based quantization suffers from some training troubles such as biased gradient or unstable gradient, rendering the encoder suboptimal. 
\item Therefore, these two quantization methods cannot ensure the latent representation ability. Expressive latent variables are significant for compression, where transmitted symbol is expected to convey more information.
\item In contrast, optimizing a compression model with additive uniform noise can be interpreted as variational optimization \cite{balle2016end} and does not encounter the training troubles. It is superior in learning an expressive latent space, as we demonstrate in this paper. 
\end{itemize}
In short, additive uniform noise is particularly well-suited to train a compression model except for the mismatch between training and test phases. Unlike the posterior and prior mismatch in VAEs \cite{kingma2016improved,dai2019diagnosing}, this mismatch originates from approximating quantization with uniform noise because the quantization error is a deterministic function regarding the signal rather than truly random noise \cite{gray1998quantization}. We thus contribute to remedy this mismatch while maintaining the advantages of additive uniform noise.

Upon our analysis, we propose a novel \emph{soft-then-hard} quantization strategy for neural image compression. Inspired by the two-stage training in recent deep generative models \cite{van2017neural,razavi2019generating,ghosh2019variational}, we first apply additive uniform noise as a \emph{soft} approximation of quantization to learn a powerful encoder. To close the mismatch caused by the noise-relaxed quantization, we then conduct ex-post tuning for the decoder with \emph{hard} quantization. It allows the decoder to be optimized for the true rate-distortion trade-off without hurting the latent representation ability. We call such proposed technique \emph{soft-then-hard} quantization strategy.

In addition, we propose to use scaled additive uniform noise by deriving a new variational upper bound on actual rate. While previous work is inflexible to control the granularity of quantization, this scaled uniform noise enables the compression model to determine element-wise adaptive quantization step. It reinforces the \emph{soft} noise-relaxed quantization and can be extended to the \emph{hard} tuning stage. As we will show, the commonly used standard uniform noise is a special case of our proposed scaled uniform noise. 

The \emph{soft-then-hard} strategy along with the scaled uniform noise is \emph{plug-and-play} to all previous noise-relaxed compression models. Experiments demonstrate that they improve the rate-distortion performance upon different base models \cite{minnen2018joint,cheng2020learned,guo20203}. Specifically, our new techniques achieve 8.9\% BD-rate savings when deployed in \cite{cheng2020learned}.

\section{Learned Lossy Image Compression} \label{section2}

From the view of classical transform coding \cite{goyal2001theoretical}, prevalent end-to-end optimized lossy image compression framework commonly follows a pipeline consisting of non-linear transform, quantization and lossless compression. Specifically, a natural image $\boldsymbol{x}$ is first mapped to latent representations $\boldsymbol{y}$, which are then quantized, yielding discrete $\boldsymbol{\hat{y}}$. Since the gradient of quantization is zero almost everywhere, it hinders the back propagation of gradients to the encoder and thus requires differentiable approximations. 

\subsection{Variational Lossy Image Compression} \label{section2.1}
Most neural image compression methods implement additive uniform noise during training to approximate the test-time quantization. Early works \cite{balle2016end,balle2018variational} illustrate the relationship between the rate-distortion objective and variational inference in this noise-relaxed case: 
\begin{equation}
\begin{aligned}
    \mathbb{E}_{\boldsymbol{x}\sim p_{\boldsymbol{x}}} D_{\rm{KL}}(q(\boldsymbol{\tilde{y}}|\boldsymbol{x})|p(\boldsymbol{\tilde{y}}|\boldsymbol{x}))  = \mathbb{E}_{\boldsymbol{x}\sim p_{\boldsymbol{x}}}&\log p(\boldsymbol{x}) + \\
    \mathbb{E}_{\bm{x}\sim p_{\bm{x}}} \mathbb{E}_{\bm{\tilde{y}}\sim q_{\bm{\tilde{y}}|\bm{x}}} [\log q(\bm{\tilde{y}}|\bm{x}) -   \log  p(\bm{x} & | \bm{\tilde{y}})-\log p (\bm{\tilde{y}})]. 
\end{aligned}
\label{equation1}
\end{equation}
The first RHS term is the log likelihood of natural images, which is a constant during optimization since the image $x$ is given in the task of compression. The second RHS term evaluates to zero in the case of additive standard uniform noise (as a stand-in for quantization during training): 
\begin{equation}
    q(\boldsymbol{\tilde{y}}|\boldsymbol{x})=q(\boldsymbol{\tilde{y}}|\boldsymbol{y})= \mathcal{U}(\boldsymbol{\tilde{y}}|\boldsymbol{y}-0.5, \boldsymbol{y}+ 0.5)=1.
\label{equation2}
\end{equation}
The rest two terms $-\log  p(\boldsymbol{x} |\boldsymbol{\tilde{y}})$ and $-\log p(\boldsymbol{\tilde{y}})$ in Eq.\ref{equation1} correspond to the weighted \textit{distortion} and the estimated \textit{rate}, respectively. The specific form of distortion is linked to the assumption on $\boldsymbol{x}$, \egno, a squared error loss equals to choosing a Gaussian assumption. But we can generalize Eq.\ref{equation1} to other distortion metrics.
Note that the actual rate is the discrete entropy of $\boldsymbol{\hat{y}}$ (at test time, $\boldsymbol{\hat{y}} = \lceil \boldsymbol{y} \rfloor$) that is non-differentiable. Following \cite{theis2017lossy}, such discrete entropy is upper-bounded by the differential entropy of $\boldsymbol{\tilde{y}}$ during training with Jensen's inequality as \footnote{We extend the derivation in \cite{theis2017lossy} with additive uniform noise ($\approx$ is got from a high-rate assumption). \cite{balle2016end} provide a statistical explanation for this inequality.}:
\begin{equation}
\begin{aligned}
    \mathbb{E}_{\boldsymbol{y}\sim q}[-\log P(\boldsymbol{\hat{y}})]  & \approx \mathbb{E}_{\boldsymbol{y}\sim q} [-\log \int_{[-0.5, 0.5]}p(\boldsymbol{y}+\boldsymbol{u})d\boldsymbol{u}] \\
    & \leq \mathbb{E}_{\boldsymbol{y}\sim q} [-\int_{[-0.5, 0.5]}\log p(\boldsymbol{y}+\boldsymbol{u})d\boldsymbol{u}] \\
    & = \mathbb{E}_{\boldsymbol{\tilde{y}}\sim q}[-\log p_{\boldsymbol{\tilde{y}}} (\boldsymbol{\tilde{y}})].
\end{aligned}
\label{equation3}
\end{equation}
Therefore, minimizing the relaxed differential entropy with distortion is equivalent to minimizing the \textit{upper bound} of the actual rate-distortion value. And the rate-distortion optimization is associated with the goal of variational inference (the LHS term in Eq.\ref{equation1}). As shown later in Section \ref{section4.2}, the standard additive uniform noise here is a special case of our derived scaled uniform noise. 

Many neural image compression approaches are built upon this variational compression framework, some of which aim to improve the entropy model \cite{minnen2018joint,lee2019context,cheng2020learned}. All of them apply additive standard uniform noise during training as a soft approximation to hard quantization. At test time, they directly quantize the latents and transmit them with entropy coding algorithms such as arithmetic coding \cite{witten1987arithmetic}. 
Therefore, there is a mismatch between training and test phases, which can be theoretically attributed to the variational relaxation of actual rate, leading to the suboptimal rate-distortion performance. It is unclear how much this mismatch is hurting performance \cite{agustsson2020universally}.

\subsection{Other Quantization Methods}

As introduced in Section \ref{section2.1}, training the lossy image compression model with additive uniform noise approximates the quantization error variationally. 
In addition, there are two other methods that tackle the non-differentiable issue of quantization in neural image compression. We briefly review them as follows. Note that we focus on the integer quantization and omit the binary quantization in some early works \cite{toderici2015variable,toderici2017full}.

\textbf{Straight-Through Estimator. }
Straight-through estimator (STE) \cite{bengio2013estimating} applies the identity gradients to pass through the hard rounding layer to enable back propagation. A few previous works adopt this method as a trivial replacement of rounding to train a generative model, such as VQ-VAE families \cite{van2017neural,razavi2019generating} and integer flow model \cite{hoogeboom2019integer}. Some compression works apply hard rounding in the forward pass, but instead use modified gradient in the backward pass \cite{theis2017lossy,mentzer2018conditional}, which we regard as variants of STE.
Since the backward and forward passes do not match, the coarse gradient before the quantization layer is certainly not the gradient of loss function. Therefore, taking the biased gradient to update the network brings some underlying problems such as unstable convergence near certain local minima, especially with improper choices of training strategy \cite{yin2019understanding}. 

\textbf{Soft-to-Hard Annealing. }
Recently, some annealing-based algorithms are proposed to approximate quantization \cite{agustsson2017soft,yang2020improving,agustsson2020universally,williams2020hierarchical}. By decreasing the value of a temperature coefficient \cite{jang2016categorical}, the differentiable approximation function goes towards the shape of hard rounding gradually. Therefore, despite using soft assignment \cite{agustsson2017soft} or soft simulation \cite{yang2020improving,agustsson2020universally} initially, these annealing-based quantization methods eventually solve the discrepancy between training and test phases when the temperature is close to zero. However, how to adjust the temperature from soft to hard is empirically determined. As a result, they suffer from fragile training. One latest work observes that annealing-based quantization achieves similar performance compared with STE when applied into integer discrete flow \cite{berg2021idf}.

\begin{figure*}[t]
 \centering
\begin{minipage}{0.24\linewidth}
 \begin{subfigure}{0.48\textwidth}
 \includegraphics[scale=0.48, clip, trim=0cm 0cm 4.3cm 0cm]{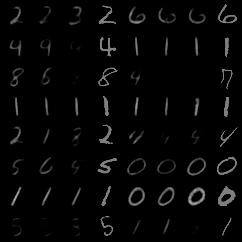}
 \end{subfigure}
\hspace{0.001\textwidth}
  \begin{subfigure}{0.48\textwidth}
 \includegraphics[scale=0.48, clip, trim=4.3cm 0cm 0cm 0cm]{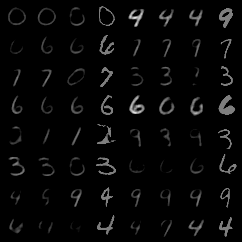}
 \end{subfigure}
\end{minipage}
\hspace{0.02\linewidth}
\begin{minipage}{0.73\linewidth}
 \begin{subfigure}{0.24\linewidth}
 \includegraphics[scale=0.38, clip, trim=1.2cm 0.8cm 1cm 1cm]{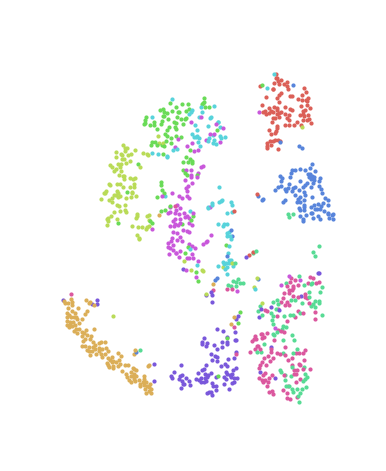}
\caption{ \label{figure1a}}
 \end{subfigure}
 \begin{subfigure}{0.24\linewidth}
 \includegraphics[scale=0.38, clip, trim=1.2cm 0.8cm 1cm 1cm]{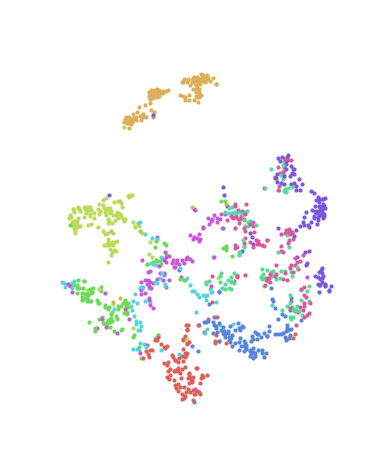}
\caption{ \label{figure1b}}
\end{subfigure}
 \begin{subfigure}{0.24\linewidth}
 \includegraphics[scale=0.38, clip, trim=1.2cm 0.8cm 1cm 1cm]{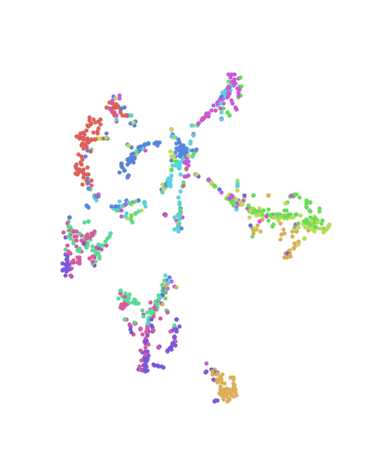}
\caption{ \label{figure1c}}
\end{subfigure}
 \begin{subfigure}{0.24\linewidth}
 \includegraphics[scale=0.38, clip, trim=1.2cm 0.8cm 1cm 1cm]{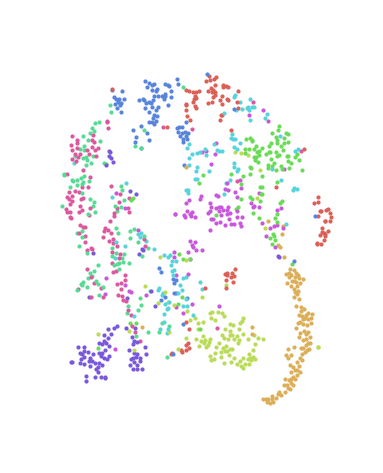}
\caption{ \label{figure1d}}
\end{subfigure}
\end{minipage}

\vspace{-0.1cm}
 \caption{Left pictures: visualizations of reconstructions, where from left to right are the reconstruction results from models trained with AUN, STE, SGA and the ground truth image. (a)-(c) The learned latent distributions by AUN, STE, SGA. (d) The learned latent distribution by tuning the AUN-pretrained model with SGA. Every color represents a category of number in the MNIST dataset.}
\label{figure1}
\end{figure*}

\section{Analysis of Quantization} \label{section3}

Unlike additive uniform noise, quantization with straight-through estimator or soft-to-hard annealing can keep training and test phases consistent because they are eventually optimized for the actual rate-distortion objective. In this section, after we investigate these three quantization methods in detail, we demonstrate that STE-based or even annealing-based quantization deteriorates the latent representation ability. An expressive latent space is extremely important in the task of compression since the transmitted symbols are always expected to convey more effective information.

\subsection{Illustrative Task} \label{section3.1}

\newcommand{\cmark}{\ding{51}}%
\newcommand{\xmark}{\ding{55}}%

\begin{table*}[t]
\centering
\caption{Comparisons of three quantization methods. Improper training strategy may cause unstable convergence of STE-based model \cite{yin2019understanding}, thus represented as -. Our proposed soft-then-hard (STH) strategy and scaled uniform noise (SUN) are meaningful.}
\label{table1}
\small
\resizebox{0.85\textwidth}{!}{
\renewcommand{\arraystretch}{1.2}
\begin{tabular}{|c|c|c|c|c|c|}
    \hline
      & AUN & STE & Annealing-Based & STH (Ours) & STH + SUN (Ours) \\
    \hline
	Train-Test Consistency &  \xmark & \cmark & \cmark & \cmark & \cmark\\
    \hline
	Latent Expressiveness &  \cmark & \xmark & \xmark & \cmark & \cmark\\
    \hline
    Variational Compression & \cmark & \xmark & \xmark & \cmark & \cmark (more flexible)\\
    \hline
    Exact Gradient & \cmark & \xmark & \cmark & \cmark & \cmark\\
    \hline
     Stable Training & \cmark & - & \xmark & \cmark & \cmark\\
    \hline
\end{tabular}}
\end{table*}

We start with an illustrative example to investigate the latent representation ability of these three quantization methods: additive uniform noise (AUN), straight-through estimator (STE) and soft-to-hard annealing (here we adopt stochastic Gumbel annealing \cite{yang2020improving}, abbreviated as SGA). We are concerned about the situation when the latent dimensionality is rigorously restricted, because it helps us examine the latent expressiveness clearly. A simplified model is used to \emph{compress} the data from MNIST dataset (main task). 
The specific experimental settings can be found in Appendix A. 

We are interested in (\romannumeral1) the reconstruction quality, and (\romannumeral2) the distribution of latent space.
In Figure \ref{figure1}, we visualize the reconstruction results of different methods. We can observe that some numbers are reconstructed to the wrong numbers since the latent dimensionality is restricted and the latent representation ability is limited. The model trained with AUN performs the best including the reconstruction diversity and accuracy. We then show the t-SNE visualization of the latent distribution \cite{van2008visualizing}, in order to examine the effects of different quantization methods. Note that we are visualizing the continuous latents that have not been quantized. Compared with the latent space in Figure \ref{figure1a} (trained by AUN), the latent spaces in Figure \ref{figure1b} (STE) and \ref{figure1c} (SGA) are more shallow, especially Figure \ref{figure1c} which tends to collapse to a low-dimensional manifold. It demonstrates that the STE-trained or the SGA-trained model cannot cover enough latent distributions to express all probable contexts. In other words, the latent representation ability of STE-trained or SGA-trained compression model does not compete with AUN-trained model.

Furthermore, in Figure \ref{figure1d}, we visualize the latent space from a model tuned by SGA but pretrained by AUN, which follows the training strategy suggested in \cite{yang2020improving}. Although the scope of latent space is almost preserved, we can observe that the latent clusters are scattered and mixed (e.g., the red and the blue clusters), which implies the inaccurate expression of the latent variables. 

This illustrative task demonstrates that additive uniform noise is superior in learning an expressive latent space especially when the latent capacity is constrained. A similar observation is found in \cite{williams2020hierarchical}, where it is termed as mode-dropping behaviour of STE. However, we show that even the annealing-based quantization will hurt the latent expressiveness in the task of compression.

\subsection{Variational or Deterministic?}  \label{section3.2}
Theoretically, if applying additive uniform noise as an approximation of quantization, the rate-distortion objective of compression is associated with the goal of variational inference, as illustrated in Eq.\ref{equation1} \cite{balle2016end,balle2018variational}. Optimizing a compression model with additive uniform noise is thereby equal to learning a variational autoencoder \cite{kingma2013auto}. However, in the case of STE-based or annealing-based quantization, the rate-distortion optimization is not variational any more since the second RHS term in Eq.\ref{equation1} (now is $\log q(\boldsymbol{\hat{y}}|\boldsymbol{x})$) is not always zero. Training a compression model with STE or annealing degrades to optimizing a deterministic autoencoder \cite{hinton2006reducing}. 
From another view, the additive uniform noise works as a regularization term for variational training, which is beneficial to learn a smooth and expressive latent space \cite{ghosh2019variational}.

\subsection{Summary of Three Quantization Methods} \label{section3.3}

\begin{figure*}[t]
 \centering
 \begin{subfigure}{0.24\linewidth}
\includegraphics[scale=1.05, clip, trim=3.8cm 8.4cm 33cm 16.5cm]{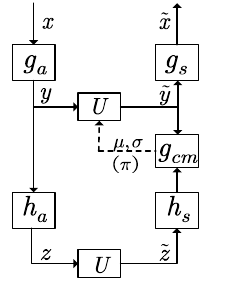}
\caption{ \label{figure2a}}
 \end{subfigure}
 \hspace{0.001\linewidth}
  \begin{subfigure}{0.24\linewidth}
\includegraphics[scale=1.05, clip, trim=8.1cm 8.4cm 28.7cm 16.5cm]{./figures/fig2/fig2.pdf}
\caption{ \label{figure2b}}
 \end{subfigure}
  \hspace{0.001\linewidth}
  \begin{subfigure}{0.24\linewidth}
\includegraphics[scale=1.05, clip, trim=12.3cm 8.4cm 24.4cm 16.5cm]{./figures/fig2/fig2.pdf}
\caption{ \label{figure2c}}
 \end{subfigure}
  \hspace{0.001\linewidth}
  \begin{subfigure}{0.24\linewidth}
\includegraphics[scale=1.05, clip, trim=16.6cm 8.4cm 20.1cm 16.5cm]{./figures/fig2/fig2.pdf}
\caption{ \label{figure2d}}
 \end{subfigure}
 \caption{(a) Applying additive uniform noise at the soft training stage, same as previous work \cite{minnen2018joint}. (b) Applying hard quantization at the ex-post tuning stage in our soft-then-hard strategy. (c) Flexible quantization with element-wise noise scale $\Delta$. (d) Combining our proposed two methods together. The gray boxes represent the components that are fixed at the ex-post tuning stage.}
\label{figure2}
\end{figure*}

As mentioned in Section \ref{section3.2}, training a compression model with STE or annealing degrades to optimizing a deterministic autoencoder. Lack of regularization term during training is one of the reasons for the weak latent representation ability. In addition, 
STE takes the biased gradient for optimization and results in searching in the negative direction \cite{yin2019understanding}. And soft-to-hard annealing will suffer from unstable training caused by infinite gradient when the temperature coefficient is closed to zero. Even if recent work tries to reduce the variance of gradients by calculating the expectation of gradients \cite{agustsson2020universally}, it requires to impose some assumptions and fails when bitrate is high. The issue of biased gradient or unstable gradient renders the encoder 
suboptimal, which is another reason for the inexpressive latent space.

In contrast, applying additive uniform noise (AUN) as a quantization approximation is superior in learning an expressive latent space. That is because: (\romannumeral1) The noise injection mechanism works as a regularization term to aid variational learning, which thus ensures the smoothness of latent space. (\romannumeral2) Applying AUN makes the training process stable with exact gradient backward. Consequently, the encoder is optimized properly to be powerful enough. But AUN still encounters the mismatch between training and test phases, resulting in rate-distortion performance degradation.

We summarize the properties of these three quantization methods as shown in Table \ref{table1}. In short, none of them enable the neural compression model to simultaneously achieve an expressive latent space and the train-test consistency. As a result, they cannot achieve the optimal rate-distortion performance for compression. In the following section, we will introduce our proposed \emph{soft-then-hard} (STH) strategy that is able to solve the train-test mismatch of AUN-based quantization while preserving the latent representation ability. In addition, we derive a new variational upper bound on actual rate that incorporates scaled uniform noise (SUN) for more flexible quantization.

\section{Proposed Methods}

\subsection{Soft-then-Hard Strategy} \label{section4.1}
We propose a novel \emph{soft-then-hard} (STH) strategy for neural image compression that contains a two-stage training process. By using this strategy, the compression model will first learn a powerful encoder with additive uniform noise that simulates the quantization layer in a soft manner. The train-test mismatch is then solved through ex-post tuning of decoder with hard quantization.

Figure \ref{figure2a} presents the training process of a conventional image compression model that is trained with additive uniform noise \cite{minnen2018joint}. It consists of an analysis encoder $\boldsymbol{g_a}$, a synthesis decoder $\boldsymbol{g_s}$, a hyper analysis encoder $\boldsymbol{h_a}$, a hyper synthesis decoder $\boldsymbol{h_s}$ and a context model $\boldsymbol{g_{cm}}$ that generates latent distribution parameters, \ieno, $\mu,\sigma$ in \cite{minnen2018joint} and $\pi,\mu,\sigma$ in \cite{cheng2020learned}. During training, additive uniform noise is added to both $\boldsymbol{y}$ and $\boldsymbol{z}$ for end-to-end optimization. This soft noise-relaxed quantization is denoted as $U$ in Figure \ref{figure2a}. 

We take the abovementioned process as the soft training stage, which is able to learn a powerful encoder. After obtaining a pretrained model, the encoders $\boldsymbol{g_a}$ and $\boldsymbol{h_a}$ will not participate in the second tuning stage. As Figure \ref{figure2b} shown,
we then directly quantize the latents $\boldsymbol{y}$, $\boldsymbol{z}$ to $\boldsymbol{\hat{y}}$, $\boldsymbol{\hat{z}}$ as the input data to tune the rest components of the compression model. The hard quantization here is denoted as $Q$ in Figure \ref{figure2b}. For simplicity, the non-parametric density estimation network of $\boldsymbol{\hat{z}}$ is also fixed, which is observed to have negligible influences experimentally. 

Since the encoders are fixed at this tuning stage, the learned latent variables will not be changed and thus the latent representation ability is preserved. At this stage, the mismatch issue of AUN-based quantization is solved, as now we are minimizing the exact discrete entropy along with the true distortion:
\begin{equation}
\begin{aligned}
\mathcal{L} = \mathbb{E}&_{\boldsymbol{y}\sim q} [-  \log P(\boldsymbol{\hat{y}}) -   \log  p_{\boldsymbol{x}|\boldsymbol{\hat{y}} }(\boldsymbol{x}|\boldsymbol{\hat{y}})].
\end{aligned}
\label{equation4}
\end{equation}
Actually, by detaching the decoder from the encoder, the entire ex-post tuning stage can be regarded as a joint optimization of two independent tasks:  treating $P(\boldsymbol{\hat{y}})$ as a new base distribution to optimize a reconstruction (generation) model, and learning a prior likelihood model to estimate the discrete latent distribution $P(\boldsymbol{\hat{y}})$. 

If the encoders are not fixed, the second tuning stage formally equals to train a compression model with STE, which will fail to learn expressive latent variables as discussed in Section \ref{section3}. In short, our proposed soft-then-hard strategy circumvents the trade-off between latent expressiveness and quantization mismatch. It is simple yet effective and does not require additional parameters.

\subsection{Scaled Uniform Noise} \label{section4.2}
In our proposed \textit{soft-then-hard} strategy, the compression model takes additive uniform noise to replace hard rounding at the soft training stage.
While adding standard uniform noise successfully approximates the integer quantization error and associates the rate-distortion optimization with variational inference, it is inflexible to control the granularity of quantization.
To sidestep this issue, we propose to learn the scale of uniform noise during training by deriving a new variational upper bound on actual rate. At test time, the adaptive noise scale will determine the quantization step.

As shown in Figure \ref{figure2c}, a new branch $h_{sq}$ will generate the noise scale $\Delta$ from $\boldsymbol{\tilde{z}}$. The noise scale is \textit{element-wise} adaptive that is encoded / decoded from hyperprior in advance. It enables the model to determine a consistent quantization granularity for arithmetic coding at both encoder and decoder. Then, we have $\boldsymbol{\tilde{y}}$ as the summation of $\boldsymbol{y}$ and a random scaled uniform noise in the interval $[-\frac{\Delta}{2},\frac{\Delta}{2}]$ as
\begin{equation}
\begin{aligned}
\Delta & =h_{sq}(\boldsymbol{\tilde{z}}), \\
\boldsymbol{\tilde{y}} & = \boldsymbol{y} + \boldsymbol{u}, \boldsymbol{u} \sim \mathcal{U}(\frac{\Delta}{2},\frac{\Delta}{2}), \\
q(\boldsymbol{\tilde{y}}|\boldsymbol{x}) & = q(\boldsymbol{\tilde{y}}|\boldsymbol{y}) = q(\boldsymbol{u}|\boldsymbol{y})=\frac{1}{\Delta}, \\
 p_{\boldsymbol{\tilde{y}}} (\boldsymbol{\tilde{y}})& = \int_{\boldsymbol{\tilde{y}}-\frac{\Delta}{2}}^{\boldsymbol{\tilde{y}}+\frac{\Delta}{2}}\frac{1}{\Delta}p(\boldsymbol{y})d\boldsymbol{y}.
\end{aligned}
\label{equation5}
\end{equation}
At test time, the learnable noise scale determines quantization step, generating $\boldsymbol{\hat{y}}$ from $\boldsymbol{y}$:
\begin{equation}
\begin{aligned}
\Delta & = h_{sq}(\boldsymbol{\hat{z}}), \\
\boldsymbol{\hat{y}} & =\Delta \cdot \lceil{\frac{\boldsymbol{y}}{\Delta}}\rfloor.
\end{aligned}
\label{equation6}
\end{equation}
Since the additive uniform noise here does not subject to standard uniform distribution, the derivation in Eq.\ref{equation3} should be modified. We derive a new variational upper bound on actual rate, which holds for scaled uniform noise $q(\boldsymbol{u}|\boldsymbol{y})$ as
\begin{equation}
\begin{aligned}
    & \mathbb{E}_{\boldsymbol{y}\sim q} [-\log P(\boldsymbol{\hat{y}})]  \\
	& \approx \mathbb{E}_{\boldsymbol{y}\sim q} [-\log \int_{[-\frac{\Delta}{2}, \frac{\Delta}{2}]}p(\boldsymbol{y}+\boldsymbol{u})d\boldsymbol{u}] \\
	& = \mathbb{E}_{\boldsymbol{y}\sim q} [-\log \int_{[-\frac{\Delta}{2}, \frac{\Delta}{2}]} q(\boldsymbol{u}|\boldsymbol{y}) \frac{p(\boldsymbol{y}+\boldsymbol{u})}{q(\boldsymbol{u}|\boldsymbol{y})}d\boldsymbol{u}] \\
    & \leq \mathbb{E}_{\boldsymbol{y}\sim q} [-\int_{[-\frac{\Delta}{2}, \frac{\Delta}{2}]} q(\boldsymbol{u}|\boldsymbol{y}) \log \frac{p(\boldsymbol{y}+\boldsymbol{u})}{q(\boldsymbol{u}|\boldsymbol{y})}d\boldsymbol{u}] \\
	& = \mathbb{E}_{\boldsymbol{\tilde{y}}\sim q}[-\log \frac{p_{\boldsymbol{\tilde{y}}} (\boldsymbol{\tilde{y}})}{q(\boldsymbol{u}|\boldsymbol{y})}] \\
	& = \mathbb{E}_{\boldsymbol{\tilde{y}}\sim q}[\log q(\boldsymbol{\tilde{y}}|\boldsymbol{x})-\log p_{\boldsymbol{\tilde{y}}} (\boldsymbol{\tilde{y}})]. \\
\end{aligned}
\label{equation7}
\end{equation}
Therefore, the rate-distortion optimization still conforms the goal of variational inference as Eq.\ref{equation1} shown. 
The true training objective which relates to the rate term is:
\begin{equation}
\begin{aligned}
\mathcal{L}_{rate}&  = \mathbb{E}_{\boldsymbol{\tilde{y}}\sim q} [- \log \int_{\boldsymbol{\tilde{y}}-\frac{\Delta}{2}}^{\boldsymbol{\tilde{y}}+\frac{\Delta}{2}}p(\boldsymbol{y})d\boldsymbol{y}] \\
	& = \mathbb{E}_{\boldsymbol{\tilde{y}}\sim q} [\log \frac{1}{\Delta} - \log p_{\boldsymbol{\tilde{y}}} (\boldsymbol{\tilde{y}})] \\
	& = \mathbb{E}_{\boldsymbol{\tilde{y}}\sim q}[-\log \frac{p_{\boldsymbol{\tilde{y}}} (\boldsymbol{\tilde{y}})}{q(\boldsymbol{u}|\boldsymbol{y})}] \\
	& \geq \mathbb{E}_{\boldsymbol{y}\sim q} [-\log P(\boldsymbol{\hat{y}})].
\end{aligned}
\label{equation8}
\end{equation}
Notice that the additive standard uniform noise in Eq.\ref{equation3} is a special case of our proposed scaled uniform noise. The noise scale allows the compression model to determine quantization granularity that is adaptive to image contexts. 

The proposed scaled uniform noise (SUN) here works at the soft quantization stage and can be extended to the hard tuning stage. As shown in Figure \ref{figure2d}, when conducting ex-post tuning with scaled uniform noise, the branch $h_{sq}$ that learns noise scale is also fixed. 
The latent variables are then quantized according to the scale values. Experiments demonstrate that coupling our proposed two methods improves performance by a large margin.


\subsection{Related Works and Discussions}

In the field of neural image compression, many works apply additive uniform noise and suffer from the mismatch between training and test phases. Most of them are dedicated to improving the entropy model.  After \cite{balle2018variational} propose the hierarchical entropy model, the works of \cite{minnen2018joint,lee2019context} design an autoregressive context model \cite{chen2016variational,van2016pixel} to capture local correlations among the latent variables. The latent distribution is also improved from zero-mean Gaussian scale model \cite{balle2018variational} to single Gaussian in \cite{minnen2018joint,lee2019context}, and recently, Gaussian mixture model \cite{cheng2020learned}. 

In addition to them, some works are closely related to our new ideas.
VQ-VAEs are perhaps the most successful VAE architectures for image and audio generations \cite{van2017neural,razavi2019generating}. In \cite{ghosh2019variational}, the authors argue that despite the name, VQ-VAEs are neither stochastic, nor variational, but they are deterministic. The work of \cite{ghosh2019variational} applies a regularization term to learn a meaningful latent space with ex-post density estimation. All of these works empirically optimize the models with two-stage training. Recently, VQ-VAE is deployed with the annealing-based quantization to mitigate the mode-dropping issue caused by STE \cite{williams2020hierarchical}. However, it is observed that the annealing-based method would perform similarly as STE \cite{berg2021idf}, which corresponds to our analysis in Section \ref{section3}. Our proposed soft-then-hard strategy is inspired by them, reasonably suitable for the task of image compression. 

\begin{figure*}[t]
 \centering
 \begin{subfigure}{0.49\linewidth}
 \includegraphics[scale=0.265, clip, trim=2.2cm 0.2cm 1cm 2cm]{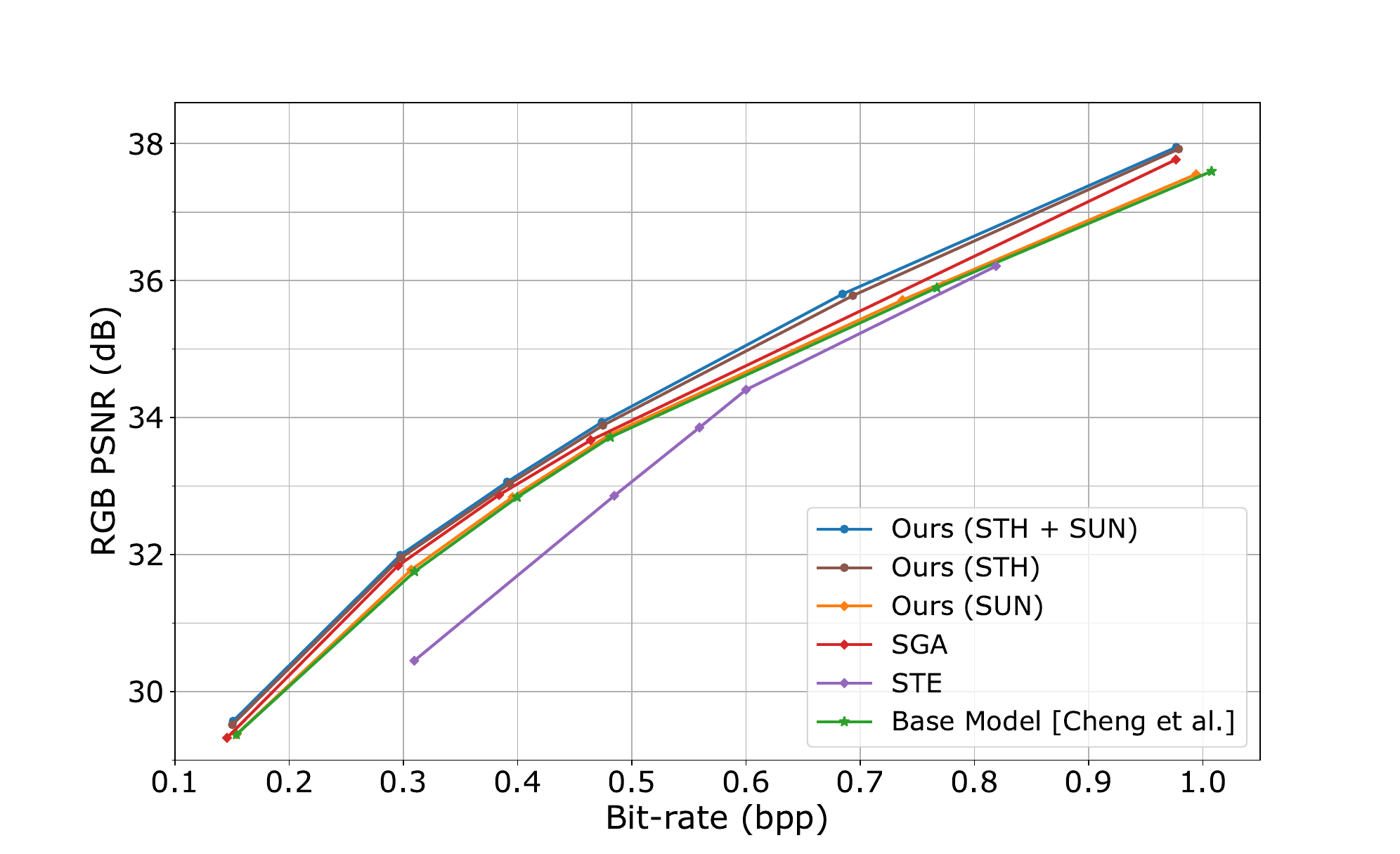}
\caption{ \label{figure3a}}
 \end{subfigure}
\hspace{0.015\linewidth}
 \begin{subfigure}{0.48\linewidth}
 \includegraphics[scale=0.265, clip, trim=2cm 0.2cm 1cm 2cm]{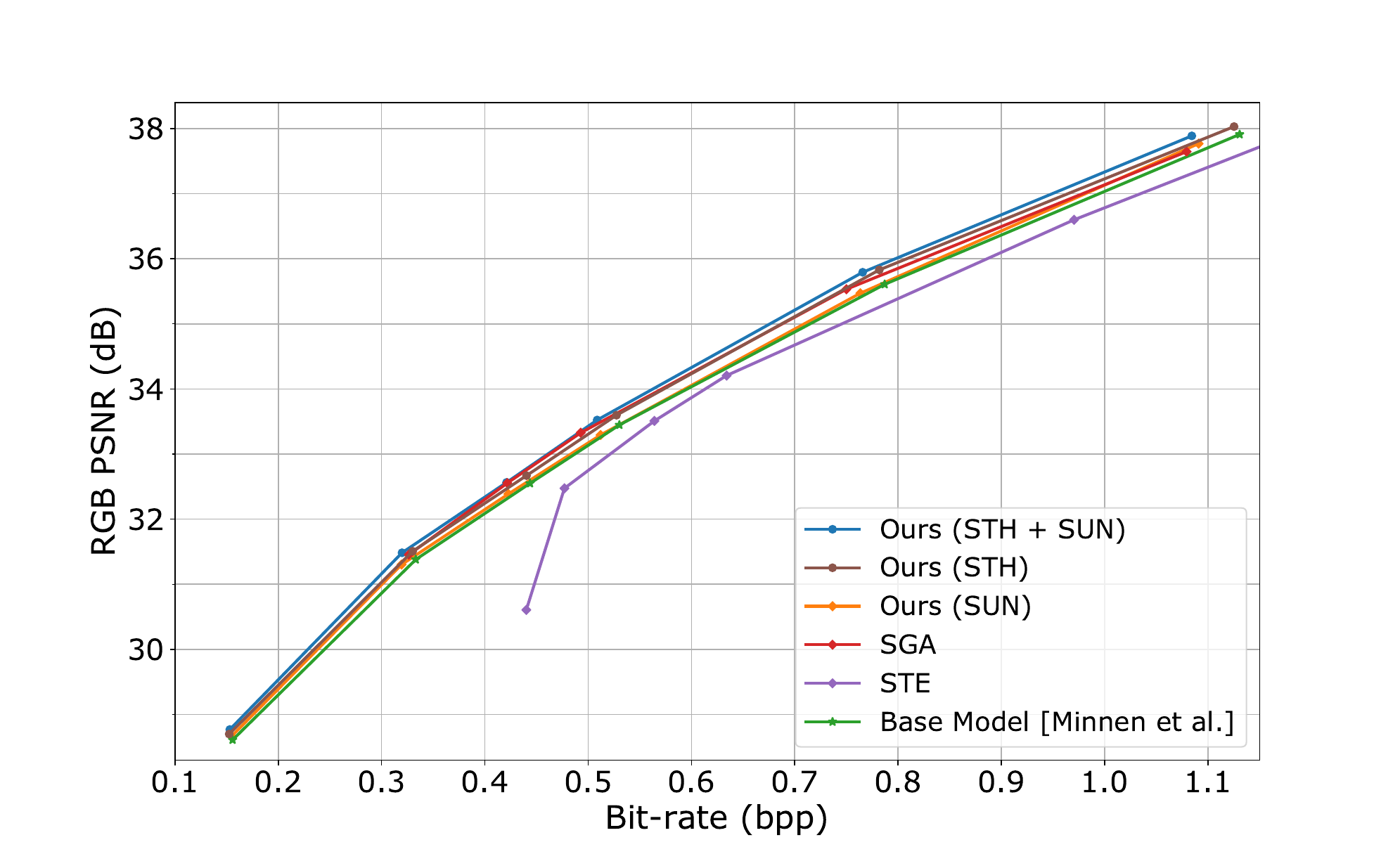}
\caption{ \label{figure3b}}
 \end{subfigure}
 \caption{Ablation results on Kodak dataset. (a) Base model is \cite{cheng2020learned}. (b) Base model is \cite{minnen2018joint}.}
\label{figure3}
\end{figure*}
As for our proposed scaled uniform noise, it is inspired by the variational dequantization in flow models \cite{ho2019flow++}. Unlike the factorized noise in  \cite{ho2019flow++}, we derive a variational upper bound on actual rate to enable flexible quantization, where the additive noise is still uniform and the model learns the noise scale. 
The work of \cite{choi2019variable} employs a group of pre-defined quantization steps with universal quantization, and builds a variable rate compression network. However, the quantization step in their work is fixed at every single bitrate that does not adapt to images. In our work, we believe that flexible quantization step is important to support spatial bit allocation, because even in traditional compression, different frequency bases are assigned with different quantization steps to adapt image contexts \cite{sullivan2012overview}. 

\section{Experiments}

Both our proposed two methods are \emph{plug-and-play}, compatible with all previous noise-relaxed lossy image compression models.
Only the new branch $h_{sq}$ that learns noise scales requires minor additional parameters. Experimentally, we observe that the mismatch issue between training and test phases deteriorates the compression performance more obviously when the compression model is more complex (see Appendix B for empirical evidence). Therefore, we evaluate our proposed new techniques upon different base models: a simplified model \cite{minnen2018joint} and two powerful models \cite{cheng2020learned,guo20203}. These models are entirely reproduced by us unless otherwise stated.
For fair comparison, we keep all experimental conditions as the same as possible (\egno, training data is not public in some previous works). 
We train the compression models on the full ImageNet training set \cite{deng2009imagenet} and test the rate-distortion performance on Kodak dataset \cite{Kodakdataset}, a widely used dataset for evaluating the performance of image compression model. Other experimental details are presented in Appendix C including the specific structures of the scale generation branch $h_{sq}$.

\subsection{Ablations}

We study the effectiveness of our proposed soft-then-hard (STH) strategy and scaled uniform noise (SUN) through extensive ablation experiments. We first compute the rate-distortion curves in terms of bit per pixel (bpp) versus peak signal-to-noise ratio (PSNR).
The loss function now is 
\begin{equation}
\begin{aligned}
\mathcal{L}&  =  \mathcal{L}_{rate} + \lambda \cdot \mathcal{L}_{distortion}\\
&  = \mathcal{L}_{rate} + \lambda \cdot \mathcal{L}_{MSE}(\boldsymbol{\hat{x}}, \boldsymbol{x}).
\end{aligned}
\label{equation9}
\end{equation}
We adjust the Lagrange Multiplier $\lambda$ from 192 to 4096 and obtain six models at multiple bitrates.

\textbf{Baseline-1.} The compression model in \cite{cheng2020learned} is a powerful codec that is trained with additive uniform noise (AUN).
As shown in Figure \ref{figure3a}, STH brings obvious gains and SUN improves the performance marginally (see Appendix D for zoom-in RD-curves). 
However, we would like to emphasize that SUN provides a novel mechanism for spatial bit allocation, which is promising for variable rate compression (\egno, using multiple noise generation branches in one model). Statistically, employing our proposed two techniques together achieves 8.9\% BD-rate savings compared with the base model. If we replace the AUN-based quantization by STE \cite{bengio2013estimating}, the performance drops a lot especially at low bitrates. 
Tuning the AUN-pretrained model by stochastic Gumbel annealing (SGA), as suggested in \cite{yang2020improving}, improves performance as well, but still has a gap to the combination of STH + SUN (even have a gap to STH alone).
Note that some points of STE and SGA are missing due to unstable training. 

\begin{figure*}[t]
 \centering
 \begin{subfigure}{0.24\linewidth}
\includegraphics[scale=0.22]{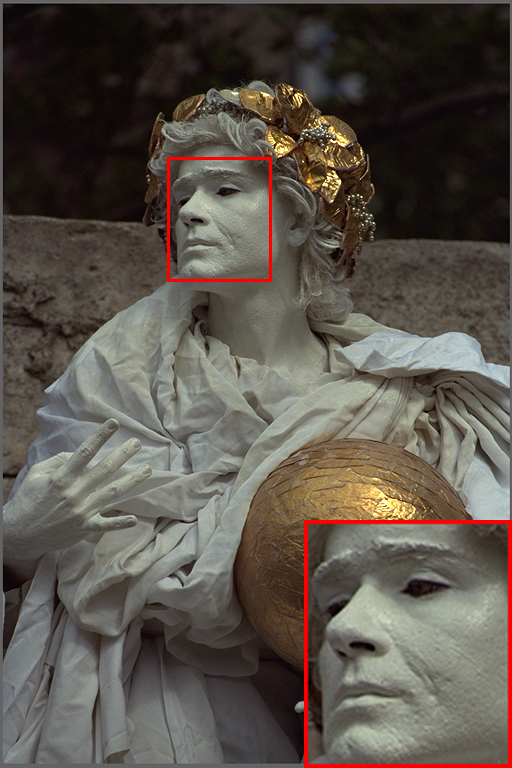}
\caption*{Ground Truth}
 \end{subfigure}
  \begin{subfigure}{0.24\linewidth}
\includegraphics[scale=0.22]{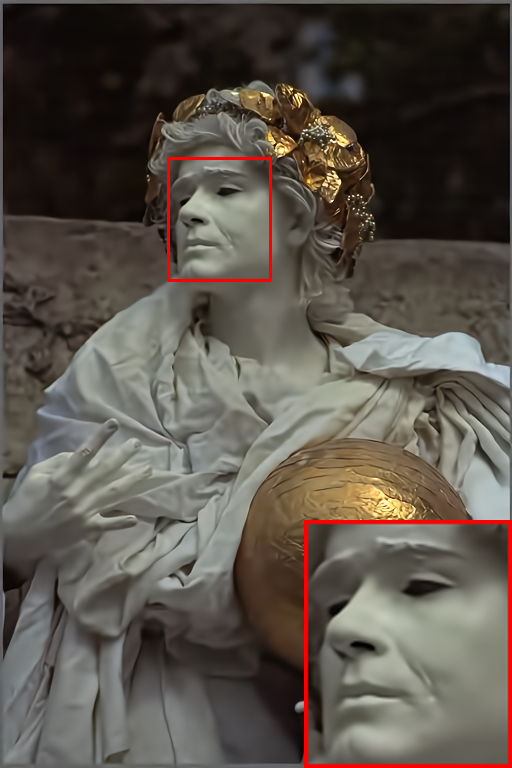}
\caption{0.210 / 32.80 / 0.9689}
 \end{subfigure}
  \begin{subfigure}{0.24\linewidth}
\includegraphics[scale=0.22]{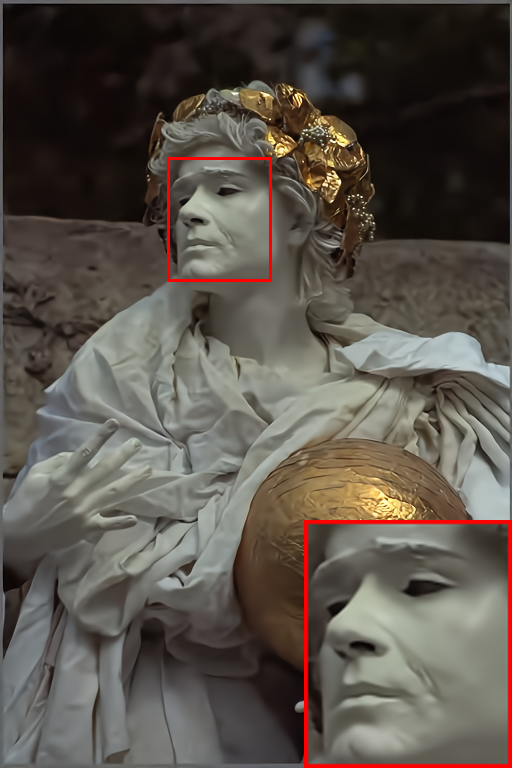}
\caption{0.203 / \textbf{33.15} / 0.9710}
 \end{subfigure}
   \begin{subfigure}{0.24\linewidth}
\includegraphics[scale=0.22]{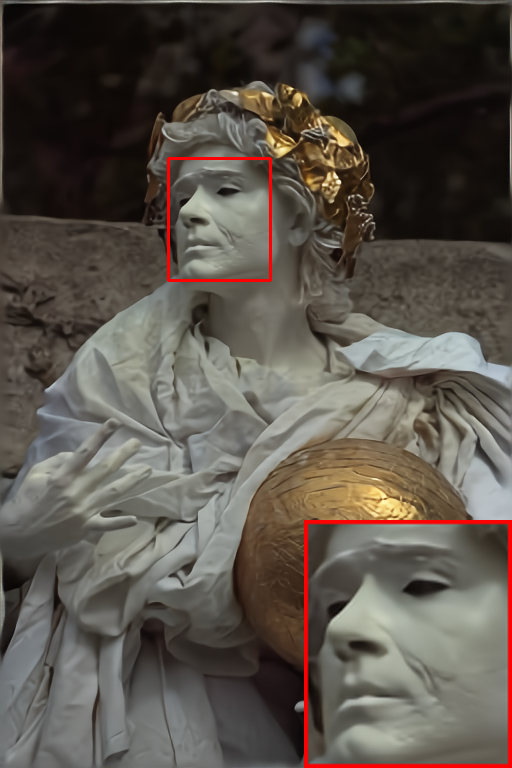}
\caption{0.169 / 28.86 / \textbf{0.9767}}
 \end{subfigure}
 \caption{Qualitative comparisons. (a) Base model \cite{cheng2020learned} optimized for PSNR. (b) Employing our methods optimized for PSNR. (c) Employing our methods optimized for MS-SSIM. The statistics are the values of bit-rate (bpp) / PSNR (dB) / MS-SSIM.}
\label{figure4}
\end{figure*}

\begin{figure*}[t]
 \centering
 \begin{subfigure}{0.22\linewidth}
\includegraphics[scale=0.125]{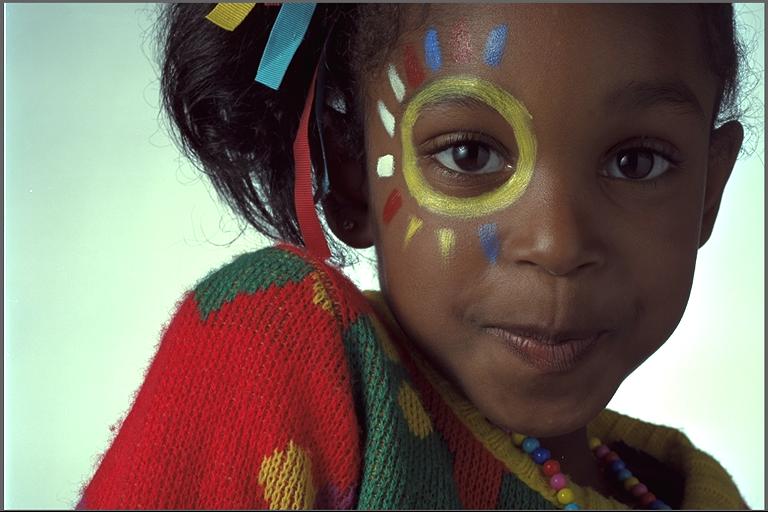}
 \end{subfigure}
 \hspace{0.01\linewidth}
  \begin{subfigure}{0.25\linewidth}
\includegraphics[scale=0.35, clip, trim=1.9cm 2.8cm 1cm 2.8cm]{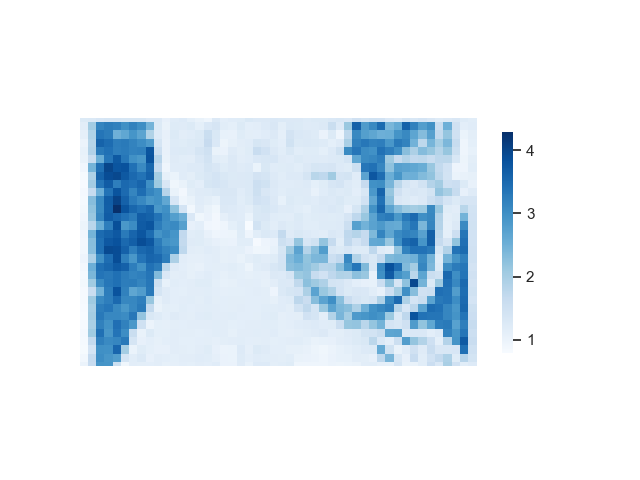}
 \end{subfigure}
  \begin{subfigure}{0.25\linewidth}
\includegraphics[scale=0.35, clip, trim=1.8cm 2.8cm 1cm 2.8cm]{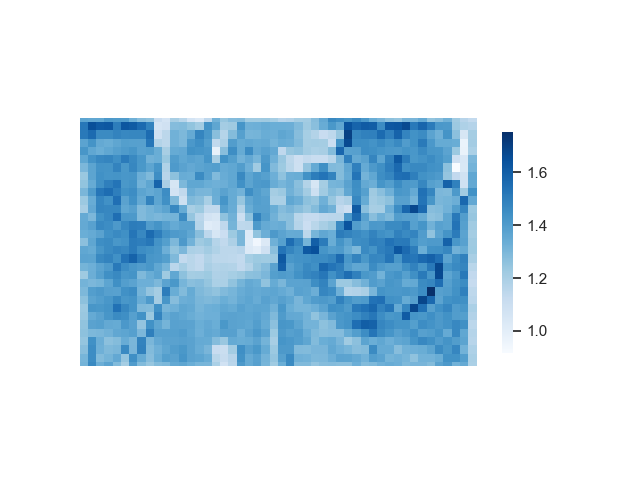}
 \end{subfigure}
   \begin{subfigure}{0.25\linewidth}
\includegraphics[scale=0.35, clip, trim=1.8cm 2.8cm 1cm 2.8cm]{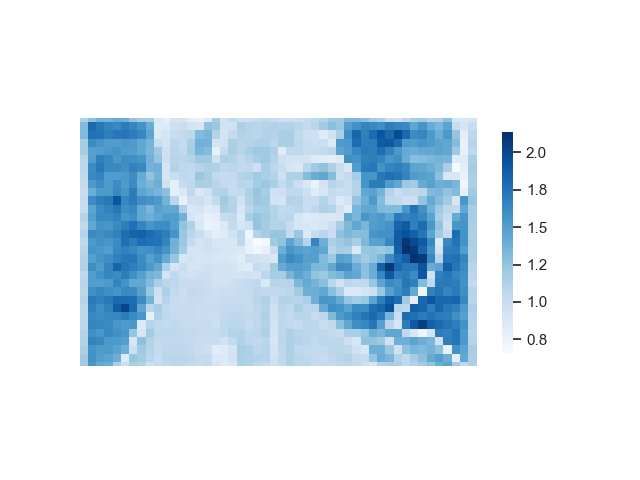}
 \end{subfigure}
\\[1pt]
  \begin{subfigure}{0.22\linewidth}
\includegraphics[scale=0.125]{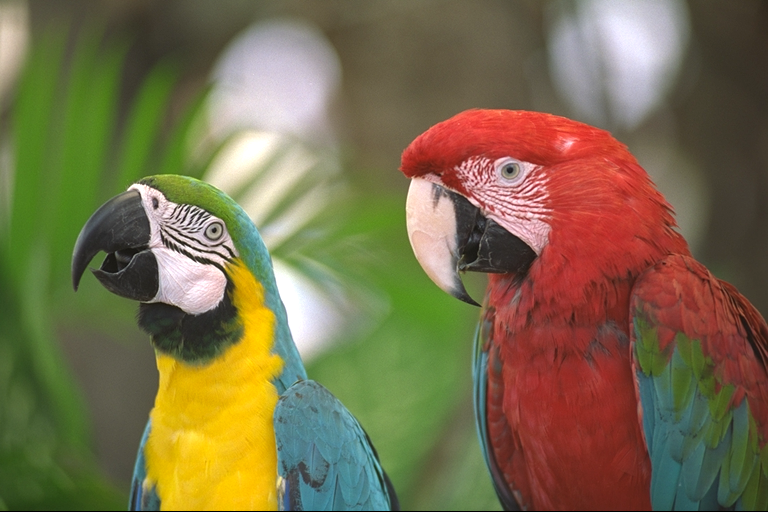}
 \end{subfigure}
 \hspace{0.01\linewidth}
  \begin{subfigure}{0.25\linewidth}
\includegraphics[scale=0.35, clip, trim=1.9cm 2.8cm 1cm 2.8cm]{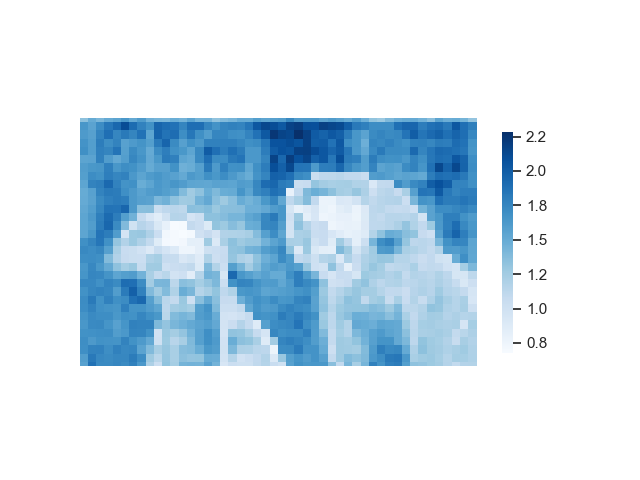}
 \end{subfigure}
  \begin{subfigure}{0.25\linewidth}
\includegraphics[scale=0.35, clip, trim=1.8cm 2.8cm 1cm 2.8cm]{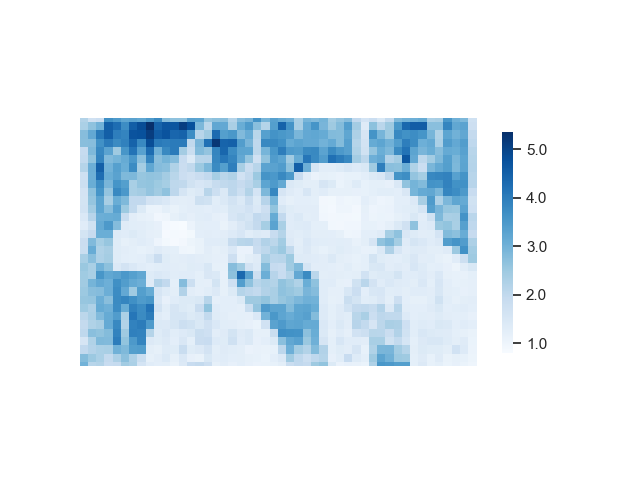}
 \end{subfigure}
   \begin{subfigure}{0.25\linewidth}
\includegraphics[scale=0.35, clip, trim=1.8cm 2.8cm 1cm 2.8cm]{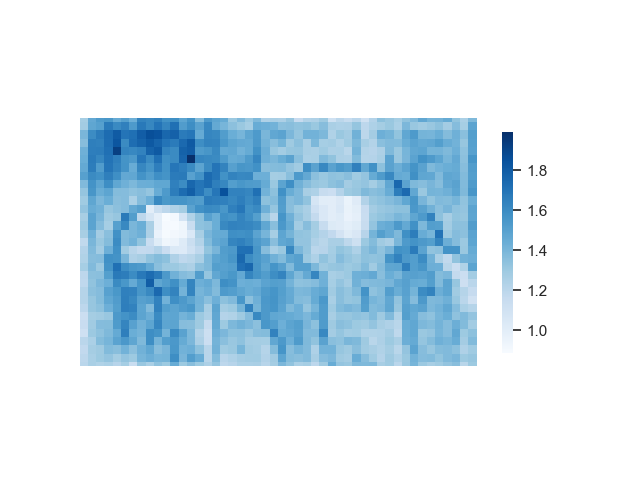}
 \end{subfigure}
 \caption{Visualizations of the noise scale. Left: ground truth. Right three columns: noise scale in different channels.}
\label{figure5}
\end{figure*}

\textbf{Baseline-2.}
Figure \ref{figure3b} presents the results upon a base model with relatively weak performance \cite{minnen2018joint}.
We find that loading the AUN-pretrained model and tuning with SGA cannot converge here. Tuning the STE-pretrained model with SGA somehow achieves good performance (the gray line) at some bitrates. This implies that SGA struggles with fragile training.
The STE-trained model encounters instability issue as well and always presents the worst rate-distortion performance. In contrast, our proposed methods, both STH and SUN, improve the performance stably. They show similar albeit weaker effects compared with the strong baseline as in Figure \ref{figure3a}.

\textbf{Other Experiments.}
We also evaluate our proposed two techniques by employing them in another powerful model \cite{guo20203}, the results of which are shown in Appendix D. Deploying our methods in such more powerful compression model delivers the state-of-the-art image compression results, outperforming VVC \cite{vvccite}, the latest compression standard. These three groups of experiments demonstrate the robustness of our proposed methods. In addition,
the soft-then-hard strategy solves the train-test mismatch and improves performance stably at all bitrates, different from \cite{agustsson2020universally}, where the annealed universal quantization would hurt RD performance at high bitrates. When optimized for MS-SSIM \cite{wang2004image}, our methods contribute stable improvements as well, which is shown in Appendix D.

\subsection{Visualizations}

\textbf{Reconstructions}. As shown in Figure \ref{figure4}, we visualize the reconstruction results when the base model is \cite{cheng2020learned}. Compared with the base model, employing our methods (STH + SUN) improves compression performance both quantitatively and qualitatively. When optimized for MS-SSIM, it achieves more pleasant perceptual quality. More visualizations are provided in Appendix E.

\textbf{Scaled Uniform Noise}. In addition, our proposed scaled uniform noise is element-wise adaptive to each latent variable. We also visualize the noise scale across different channels as presented in Figure \ref{figure5}. Notice that the learned noise scale is highly correlated to image textures and covers different contexts across channels. It verifies that our derived scaled uniform noise is effective to generate image-adaptive quantization step.

\section{Conclusion}

In this paper, we rethink the three quantization methods that are implemented as differentiable approximations for neural image compression. Among them, we demonstrate that additive uniform noise is superior to STE-based and even annealing-based quantization in terms of the latent representation ability through our detailed analysis. We propose a novel soft-then-hard quantization strategy that achieves train-test consistency and latent expressiveness simultaneously.
We also derive a new variational upper bound on actual rate that incorporates the scale of additive uniform noise into optimization and thus enable flexible quantization. Our proposed two methods are simple yet effective, achieving stable improvements at any bitrates.


\section*{Acknowledgements}
Zhibo Chen is the corresponding author. This work was supported in part by NSFC under Grant U1908209, 61632001, and the National Key Research and Development Program of China 2018AAA0101400.

\bibliography{example_paper}
\bibliographystyle{icml2021}

\newpage

\section*{Appendix A: Illustrative Task}
\label{appendix_a}
In our main paper, we conduct an illustrative experiment to show that additive uniform noise is superior in learning an expressive latent space for compression. Here we introduce the detailed settings of this experiment.

The core of this experiment is an image compression task. We build the compression model that first encodes the image from MNIST dataset to latent variables. We then try three quantization methods to discretize the latent variables for end-to-end optimization. A decoder will correspondingly generate the reconstruction from the quantized latent variables. Since the image resolution in MNIST dataset is $28\times28$, the network architecture is designed as follows:
\vspace{-0.2cm}
\begin{table}[h]
\centering
\caption{Network architecture in this illustrative task.}
\label{table2}
\renewcommand{\arraystretch}{1.2}
\begin{tabular}{|c|c|}
    \hline
      \textbf{Encoder} & \textbf{Decoder} \\
    \hline
	Conv: 5$\times$5 c32 s2 & Deconv: 7$\times$7 c32 s1 \\
	LeakyReLU & LeakyReLU \\
	Conv: 5$\times$5 c32 s2 & Deconv: 5$\times$5 c32 s2 \\
	LeakyReLU & LeakyReLU \\
	Conv: 7$\times$7 c4 s1 & Deconv: 5$\times$5 c1 s2 \\
    \hline
\end{tabular}
\end{table}
This model will transform the image to a four-dimension latent vector, \textit{i}.\textit{e}., the $28\times28\times1$ image will be mapped to the $1\times1\times4$ latent variable. Ideally, one continuous real number is able to represent any information if the transform network is very powerful. However, the encoder network here is not strong enough. We thereby design to restrict the latent capacity to investigate the latent representation ability that is learned with different quantization methods. 

The model is optimized for the rate-distortion objective. The distortion is measured by mean square error between the original image $\boldsymbol{x}$ and the reconstructed image $\boldsymbol{\hat{x}}$. The rate here is measured by the $\mathcal{L}_2$ norm of the quantized latent variables as the continuous log-likelihood $\log p(\boldsymbol{\tilde{y}})$. It is equal to assume a zero-mean Gaussian distribution with fixed scale on $\boldsymbol{\tilde{y}}$. The overall loss function is:
\begin{equation}
\begin{aligned}
\mathcal{L}&  =  \mathcal{L}_{rate} + \lambda \cdot \mathcal{L}_{distortion}\\
&  = \mathcal{L}_{2}(\boldsymbol{\tilde{y}}) + \lambda \cdot \mathcal{L}_{MSE}(\boldsymbol{\hat{x}}, \boldsymbol{x}).
\end{aligned}
\label{equation10}
\end{equation}
We set the Lagrange Multiplier $\lambda$ as 10 and use Adam optimizer with learning rate 1e-3 for optimization. We visualize the results in our main paper by selecting the best model during the total 80-epoch training process.

\section*{Appendix B: Train-Test Mismatch}
\label{appendix_b}

This section provides evidences along with some analyses to show that the mismatch between training and test phases is more serious in complex compression model. 

The train-test mismatch is measured by the performance gap between soft quantization (additive uniform noise) and hard rounding. Specifically, we can try to use additive uniform noise to test the (estimated) compression performance on Kodak dataset, although it is not a practical compression process. In Table \ref{table3}, we present the distortion gap between training and test phases that is measured by the difference between the estimated PSNR value and the true PSNR value: $\rm{Gap} = \rm{PSNR}_{soft} - \rm{PSNR}_{hard}$.
\vspace{-0.2cm}
\begin{table}[h]
\centering
\caption{Distortion mismatch between training and test phases.}
\label{table3}
\renewcommand{\arraystretch}{1.2}
\resizebox{0.98\columnwidth}{!}{
\begin{tabular}{|c|c|c|c|c|c|c|}
    \hline
     $\lambda$ & 192 & 512 & 768 & 1024 & 2048 & 4096 \\
    \hline
    Baseline-1 Gap & \textbf{0.26} & \textbf{0.33} & \textbf{0.33} & \textbf{0.28} & \textbf{0.40} & \textbf{0.50}\\
    \hline
    Baseline-2 Gap & 0.14 & 0.17 & 0.14 & 0.21 & 0.29 & 0.04\\
    \hline
\end{tabular}}
\end{table}

Baseline-1 is the model of \cite{cheng2020learned}, more powerful than Baseline-2 \cite{minnen2018joint}. We can observe that the distortion gap is more serious in the complex base model (Baseline-1). We deduce that perhaps it is due to the posterior collapse issue, since a sufficiently powerful decoder will tend to ignore the posterior in VAEs.

\begin{figure*}[t]
 \centering
 \begin{subfigure}{0.48\linewidth}
 \includegraphics[scale=0.30, clip, trim=0.5cm 0.2cm 1cm 2cm]{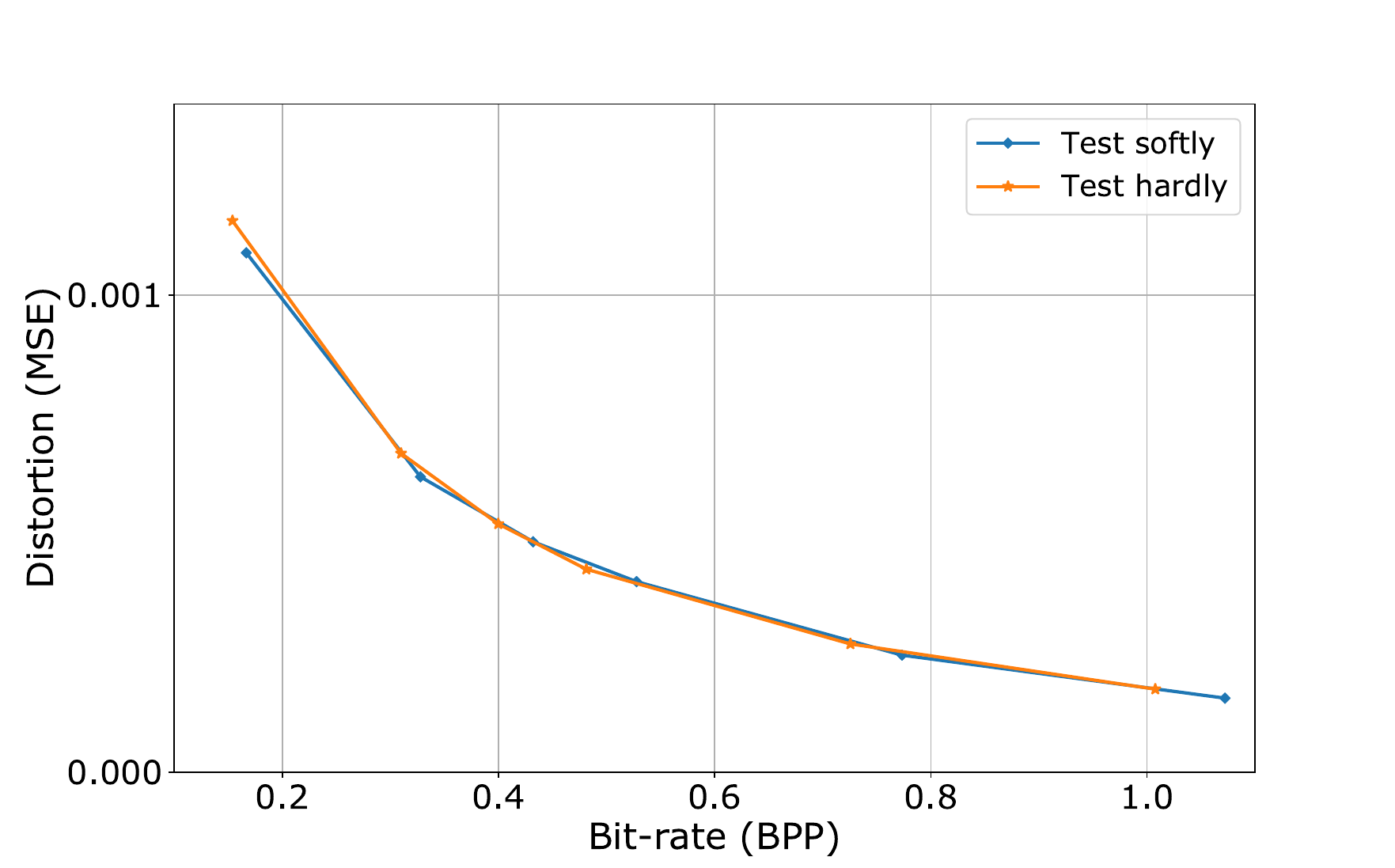}
\caption{ \label{figure6a}}
 \end{subfigure}
\hspace{0.015\linewidth}
 \begin{subfigure}{0.48\linewidth}
 \includegraphics[scale=0.30, clip, trim=0.5cm 0.2cm 1cm 2cm]{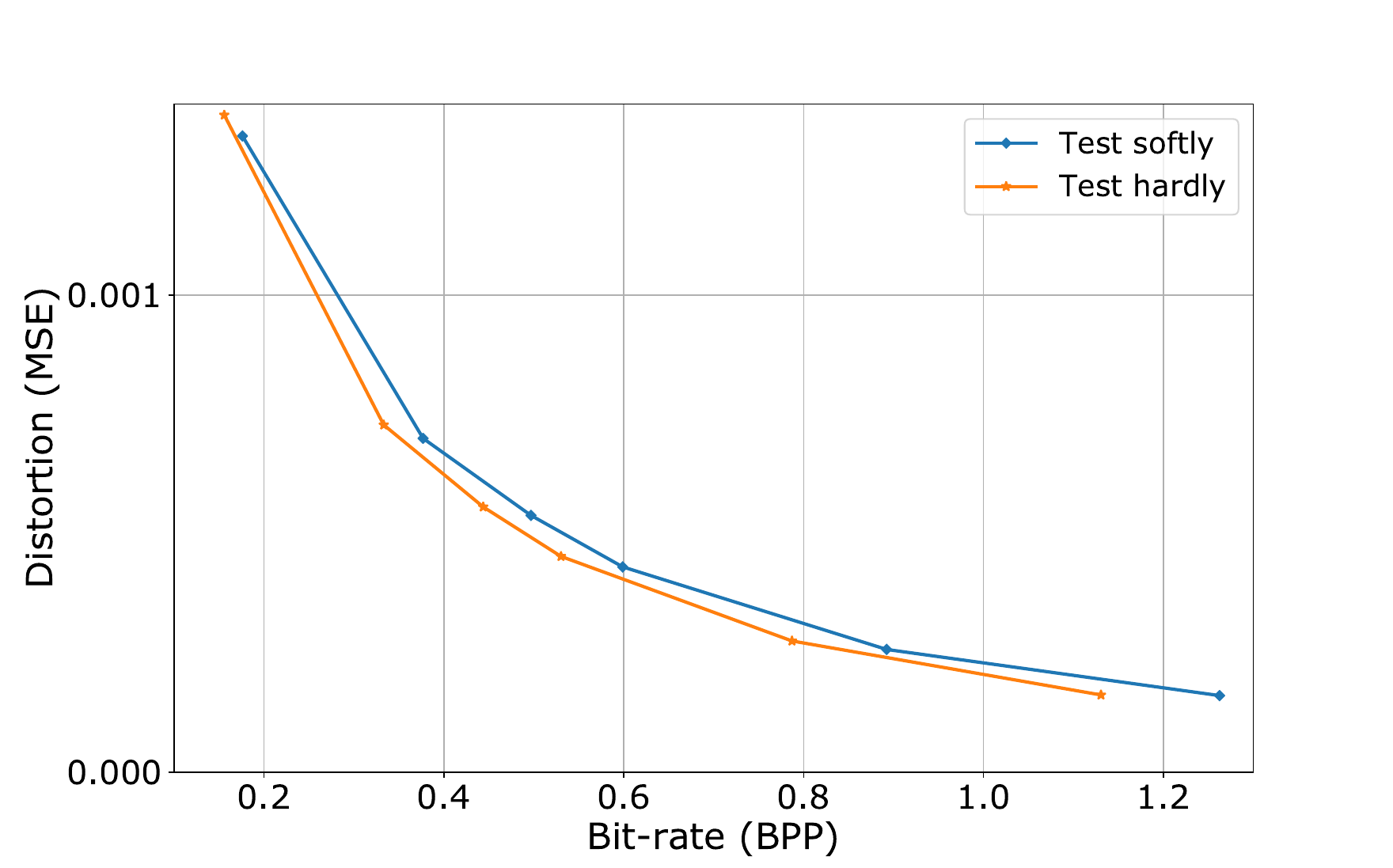}
\caption{ \label{figure6b}}
 \end{subfigure}
 \caption{The rate-distortion performance mismatch between soft quantization (additive uniform noise) and hard rounding. (a) Base model is \cite{cheng2020learned}. (b) Base model is \cite{minnen2018joint}. Evaluating on Kodak dataset.}
\label{figure6}
\end{figure*}

In addition, we draw both the estimated and the true rate-distortion (RD) curves upon these two base models as shown in Figure \ref{figure6}, \ieno, test with soft quantization (additive uniform noise) and test with hard quantization. Here we directly compute the mean square error to stand for distortion. When baseline is a complex model \cite{cheng2020learned}, the true rate-distortion curve coincides with the estimated curves as shown in Figure \ref{figure6a}. However, it would be surprising that the true RD performance of Baseline-2 is better than the estimated performance that corresponds to the soft training objective. Actually, it is reasonable because the noise-relaxed compression models are optimized to minimize the variational upper bound of actual rate. Therefore, the estimated rate is larger than the true rate. From another view, in simple compression models, the actual rate-distortion performance is better than the estimated performance. But this performance improvement (from training to test) is weakened in complex models such as \cite{cheng2020learned}, which also implies that the mismatch between training and test phases is more serious in complex compression model.

\section*{Appendix C: Ablation Settings}
\label{appendix_c}

We conduct rigorous ablation study to verify the effectiveness of our proposed techniques, as mentioned in our main paper. We here complement some experimental settings of our ablations and introduce some specific architectures.

\subsection*{Training Details}
We train the compression models on the full ImageNet dataset.
Original images are cropped to $256 \times 256$ patches. Minibatches of 8 of these patches are used to update network parameters that is trained on single RTX 2080 Ti GPU.
We apply Adam optimizer with learning rate decay strategy. At the soft training stage, the initial learning rate is 5e-5 and degrades to 1e-5 after 400,000 iterations. We obtain the pre-trained model by selecting the best model during 2,000,000 iterations that is evaluated on Kodak dataset. After accomplishing the soft training stage, we employ scaled uniform noise in the pre-trained model by finetuning the noise-generation branch with 500,000 iterations. Then we conduct ex-post tuning with hard quantization. At this stage, we finetune the decoder for 500,000 iterations as well. During the second and the third stage, the learning rate is 5e-5 and degrades to 1e-5 after 200,000 iterations. The latent channel number is increased at high bitrates to avoid bottleneck issue following \cite{balle2018variational}. Specifically, we assign M=192 channels for low or intermediate bitrates, and assign M=320 channels when $\lambda$ is $2048$ or $4096$.

\subsection*{Reproducing Details}
We investigate the effects of our methods upon three base models \cite{minnen2018joint,cheng2020learned,guo20203}. All of them are reproduced by us with Pytorch implementations. 
The network structures are reproduced as their paper reported exactly. Our reproduced performance has a gap to their reported statistics, which may be caused by the difference of training data. Specifically, we use the full ImageNet training set without extra selection. While \cite{minnen2018joint} do not mentioned the training set,  \cite{cheng2020learned} adopt the subset of ImageNet for training with coarse selection and \cite{guo20203} use some high-resolution datasets for training. Therefore, there is a gap between our reproduced results and their reported results. 

\subsection*{Structures of the New Branch}
Our proposed soft-then-hard strategy does not require additional network parameters. Our proposed another technique, the scaled uniform noise, requires a new branch to generate noise scales. Since the value of noise scale is positive, we adopt an exponential layer to activate the final output of this branch. The specific structure of this branch is shown in Figure \ref{figure7}:
\vspace{-0.2cm}
\begin{figure}[h]
 \centering
 \includegraphics[scale=0.5, clip, trim=31.2cm 16.6cm 6.6cm 2.8cm]{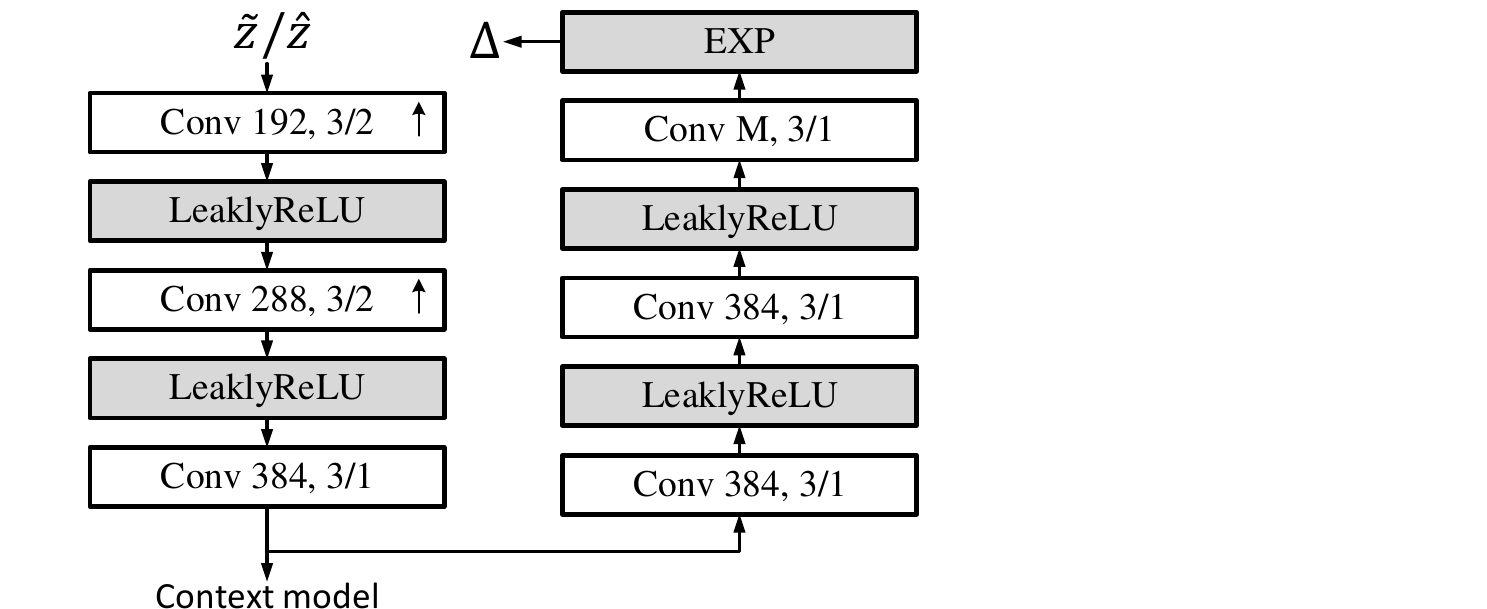}
\caption{The structure of our proposed branch that generates noise scale $\Delta$. The left several layers are shared with context model. The input is $\tilde{z}$ for soft quantization and $\hat{z}$ for hard quantization. The value of $\Delta$ is clamped to avoid extreme value. \label{figure7}}
\end{figure}

\section*{Appendix D: More Experimental Results}
\label{appendix_d}

\begin{figure*}[h]
 \centering
 \includegraphics[scale=0.55, clip, trim=0.5cm 0.2cm 1cm 2cm]{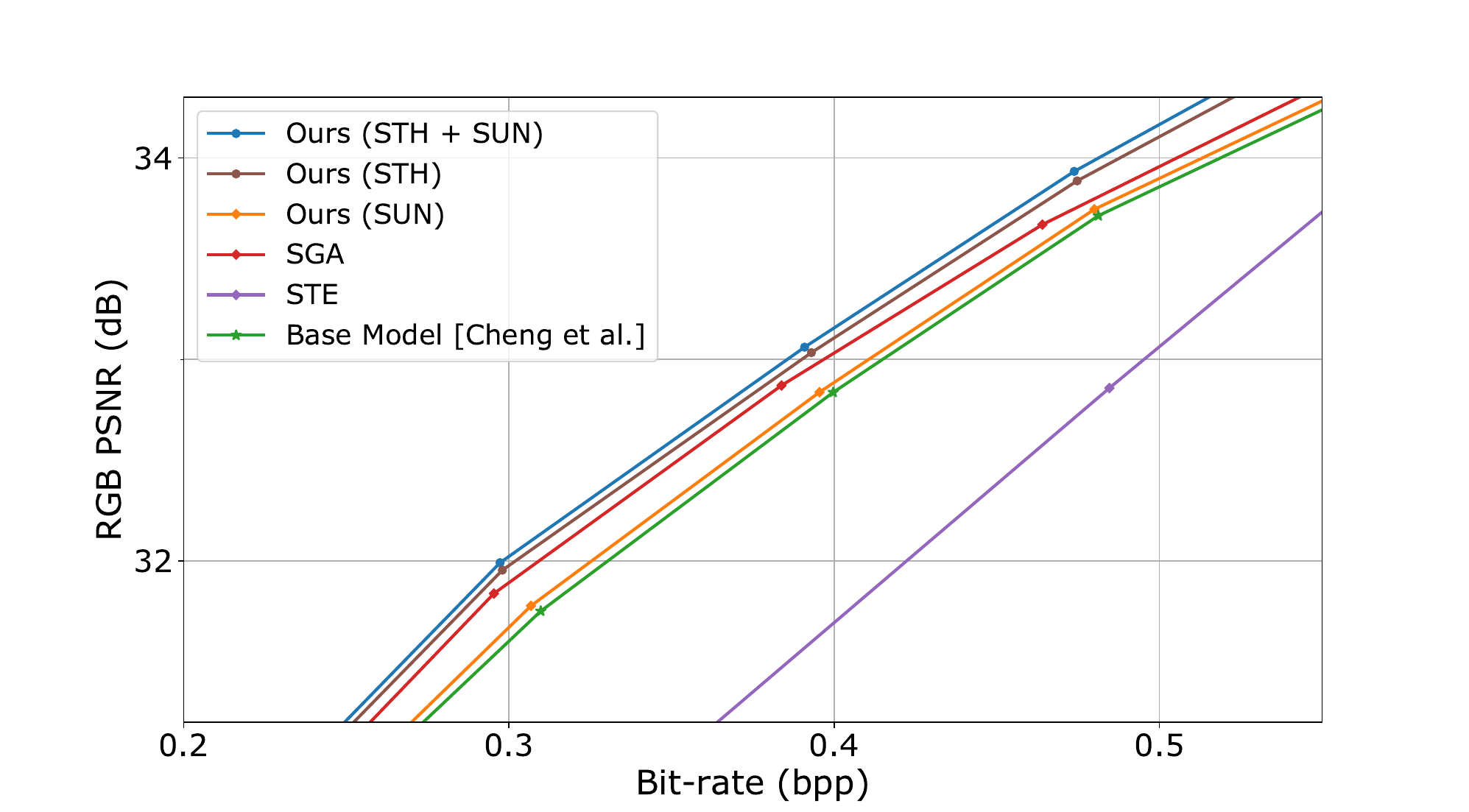}
 \caption{The zoom-in rate-distortion curve in base model \cite{cheng2020learned}. Evaluating on Kodak dataset.}
\label{figure8}
\end{figure*}

Here we first provide the zoom-in RD-curves about the ablation study in base model \cite{cheng2020learned} for better visualization. As shown in Figure \ref{figure8}, our proposed scaled uniform noise (SUN) achieves considerable improvements by around 0.1 dB at intermediate bitrates. 

\begin{figure*}[h]
 \centering
 \includegraphics[scale=0.49, clip, trim=0.5cm 0.2cm 1cm 2cm]{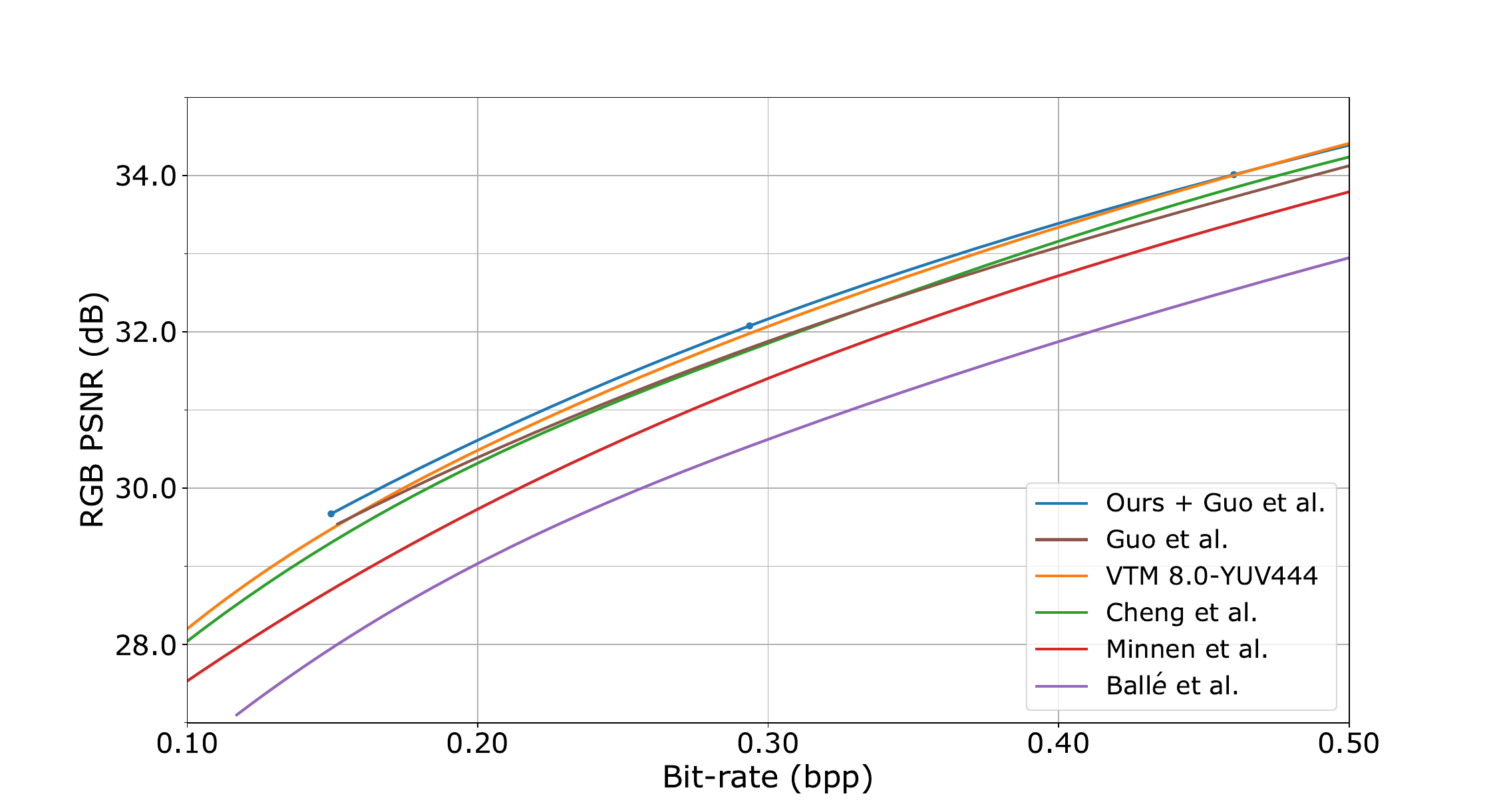}
 \caption{Employing our proposed two techniques in base model \cite{guo20203}. It helps us achieve the state-of-the-art image compression performance, outperforming all previous neural image compression approaches and the latest image compression standard.}
\label{figure9}
\end{figure*}

As shown in Figure \ref{figure9}, we also present the results of deploying our methods in base model \cite{guo20203}, which delivers the state-of-the-art image compression performance. It even outperforms the H.266/VVC standard on Kodak dataset (we use the VTM-8.0 anchor with YUV444 format and all intra mode). Here the statistics of previous works of neural image compression are taken from their report in papers including \cite{balle2018variational}, \cite{minnen2018joint}, \cite{cheng2020learned} and \cite{guo20203}.

\begin{figure*}[h]
 \centering
 \includegraphics[scale=0.49, clip, trim=0.5cm 0.2cm 1cm 2cm]{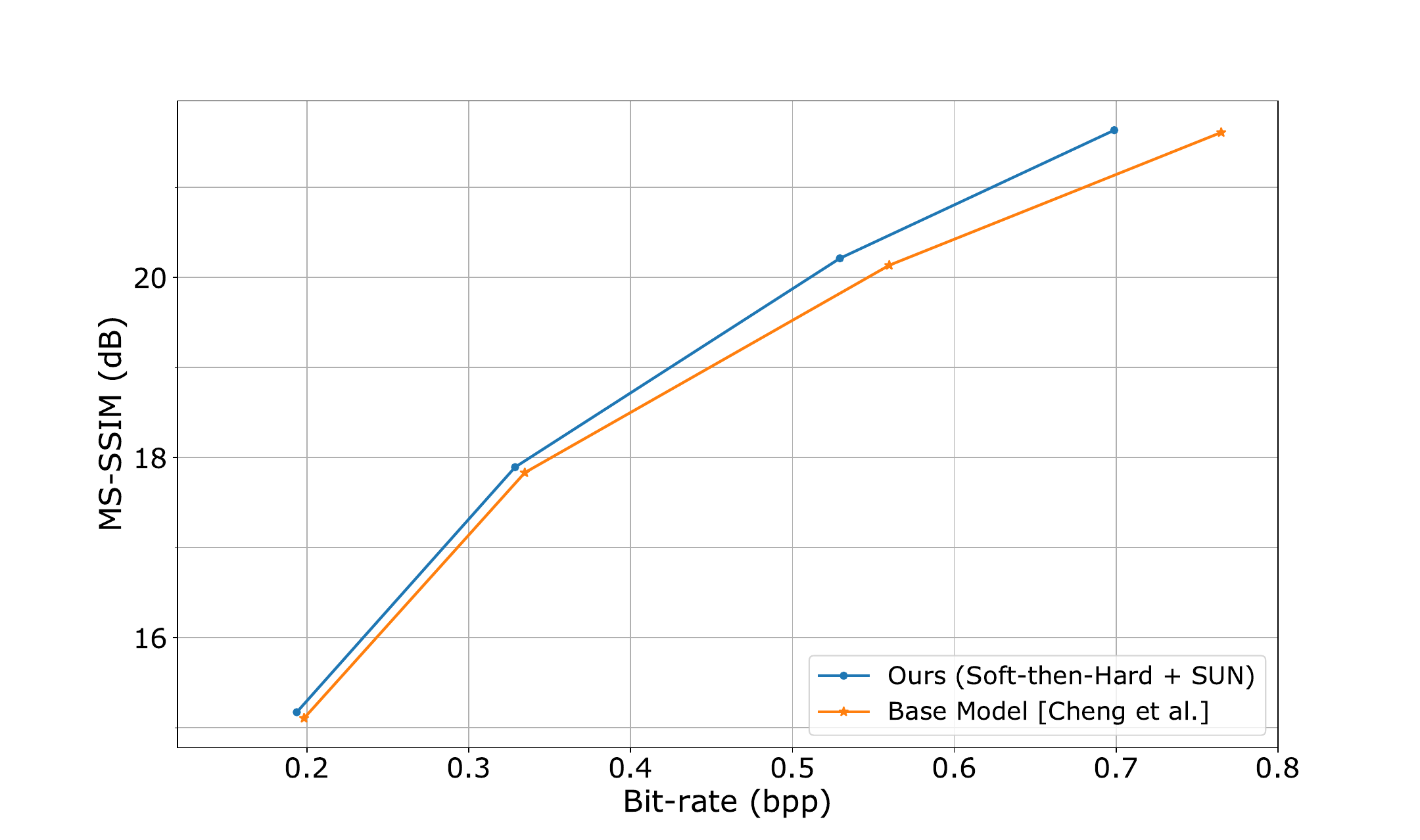}
 \caption{Ablation results in base model \cite{cheng2020learned}. Optimized for MS-SSIM.}
\label{figure10}
\end{figure*}

Another important comparison is about the MS-SSIM-optimized case, the metric of which is more consistent with human perceptual quality \cite{wang2004image}. The loss function now is:
\begin{equation}
\begin{aligned}
\mathcal{L}&  =  \mathcal{L}_{rate} + \lambda \cdot \mathcal{L}_{distortion}\\
&  = \mathcal{L}_{rate} + \lambda \cdot (1 - \mathcal{L}_{MS-SSIM}(\boldsymbol{\hat{x}}, \boldsymbol{x})).
\end{aligned}
\label{equation11}
\end{equation}
We train models at four different bitrates with $\lambda=16,40,100,180$ (latent channel number M=320 when $\lambda=100$ or $180$). Our methods also bring obvious gains in base model \cite{cheng2020learned} as shown in Figure \ref{figure10}. 

In summary, our proposed new methods are robust to bring stable improvements of rate-distortion performance at any bitrate in different base models.

\section*{Appendix E: More Visualizations}
\label{appendix_e}

\begin{figure*}[t]
 \centering
 \begin{subfigure}{0.33\linewidth}
\includegraphics[scale=0.21]{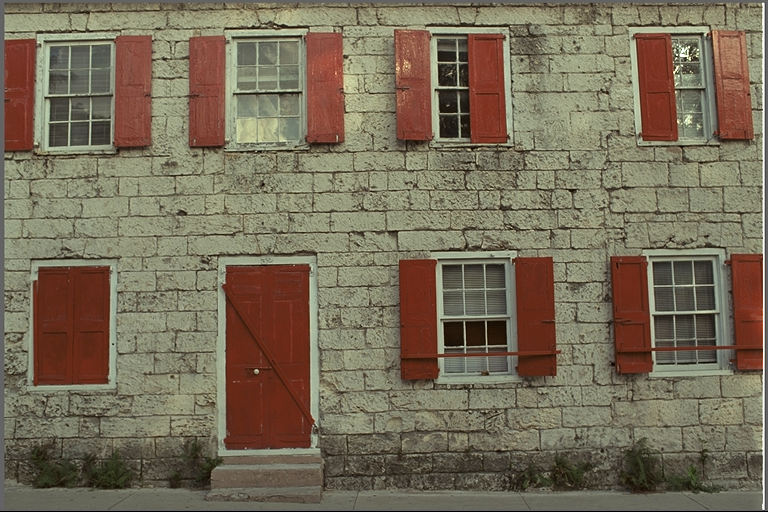}
\caption*{Ground Truth}
 \end{subfigure}
  \begin{subfigure}{0.33\linewidth}
\includegraphics[scale=0.21]{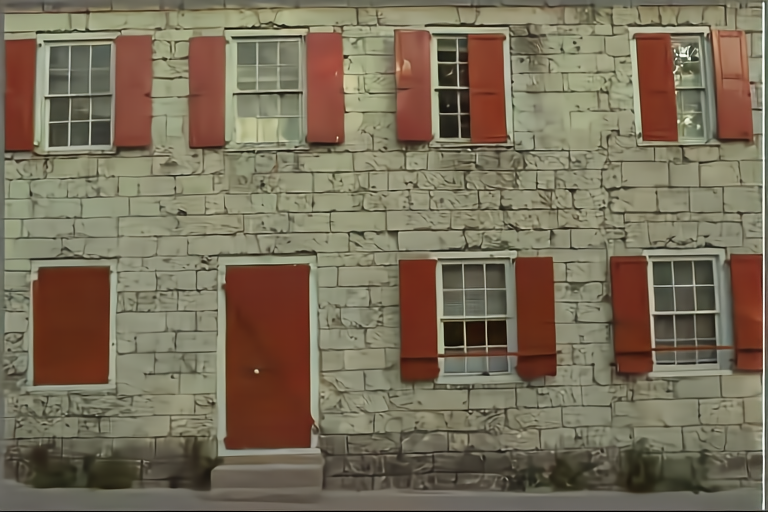}
\caption*{(a) 0.238bpp / 26.71dB / 0.9390}
 \end{subfigure}
  \begin{subfigure}{0.33\linewidth}
\includegraphics[scale=0.21]{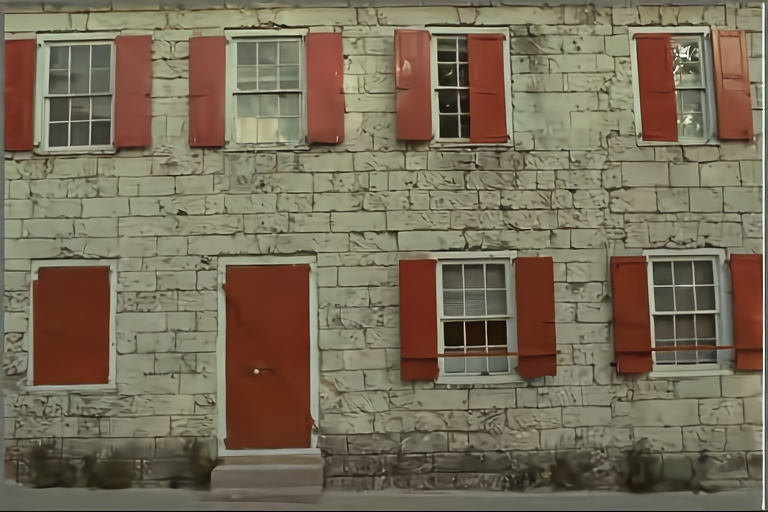}
\caption*{(b) 0.232bpp / \textbf{26.84}dB / 0.9411}
 \end{subfigure}

 \begin{subfigure}{0.33\linewidth}
\includegraphics[scale=0.21]{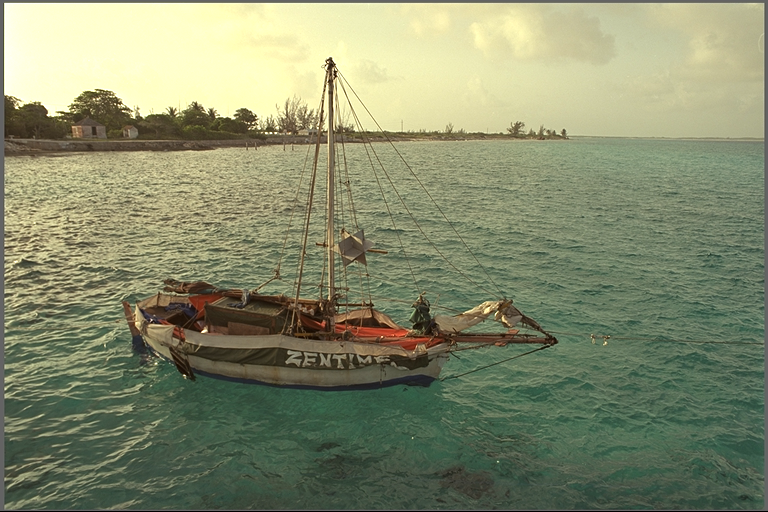}
\caption*{Ground Truth}
 \end{subfigure}
  \begin{subfigure}{0.33\linewidth}
\includegraphics[scale=0.21]{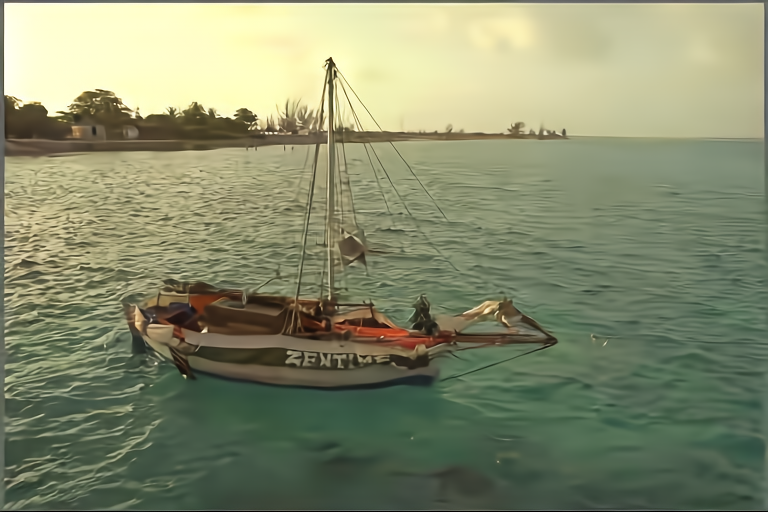}
\caption*{(a) 0.179bpp / 27.98dB / 0.9227}
 \end{subfigure}
  \begin{subfigure}{0.33\linewidth}
\includegraphics[scale=0.21]{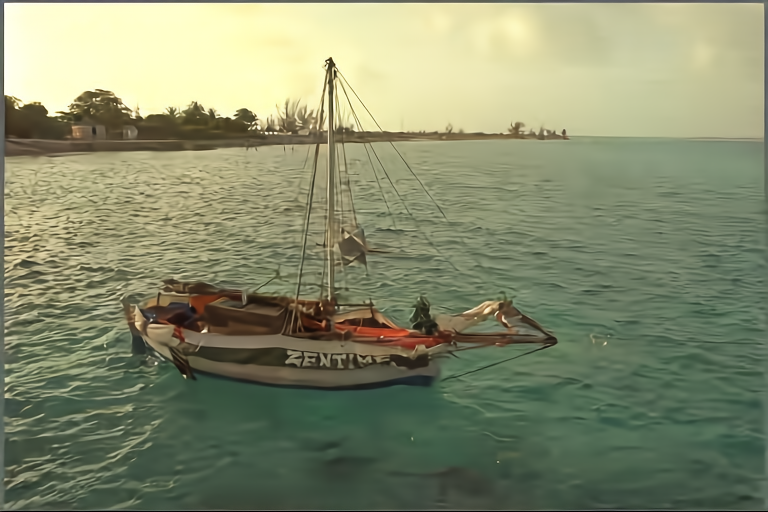}
\caption*{(b) 0.179bpp / \textbf{28.17}dB / 0.9269}
 \end{subfigure}
 \caption{Visual comparisons. (a) Base model \cite{cheng2020learned} optimized for PSNR. (b) Employing our methods in this base model optimized for PSNR. The statistics are the values of bit-rate (bpp) / PSNR (dB) / MS-SSIM.}
\label{figure11}
\end{figure*}

\begin{figure*}[t]
 \centering
 \begin{subfigure}{0.33\linewidth}
\includegraphics[scale=0.21]{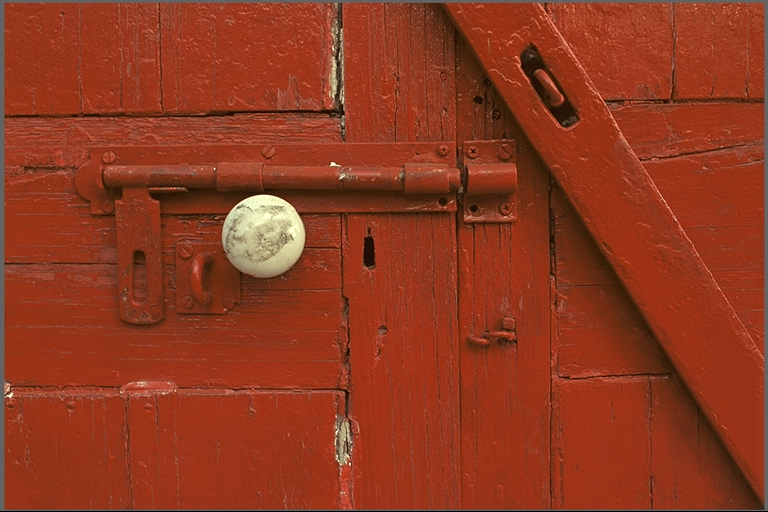}
\caption*{Ground Truth}
 \end{subfigure}
  \begin{subfigure}{0.33\linewidth}
\includegraphics[scale=0.21]{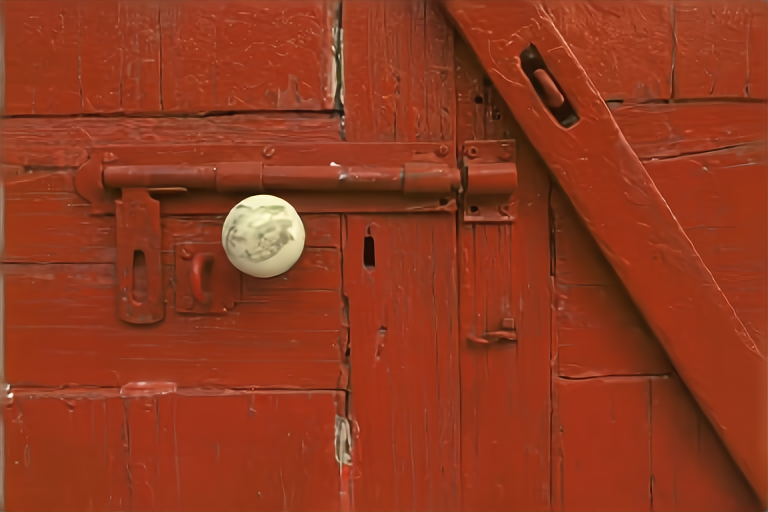}
\caption*{(a) 0.187bpp / 28.24dB / 0.9614}
 \end{subfigure}
  \begin{subfigure}{0.33\linewidth}
\includegraphics[scale=0.21]{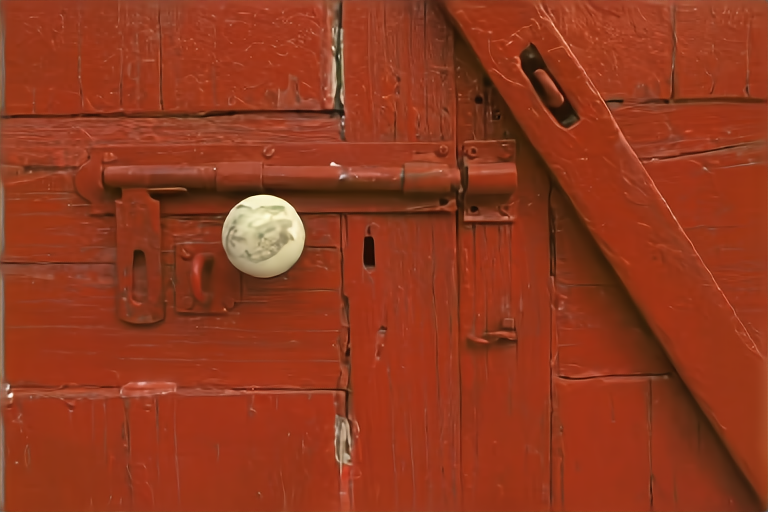}
\caption*{(b) 0.184bpp / 28.29dB / \textbf{0.9623}}
 \end{subfigure}

 \begin{subfigure}{0.33\linewidth}
\includegraphics[scale=0.21]{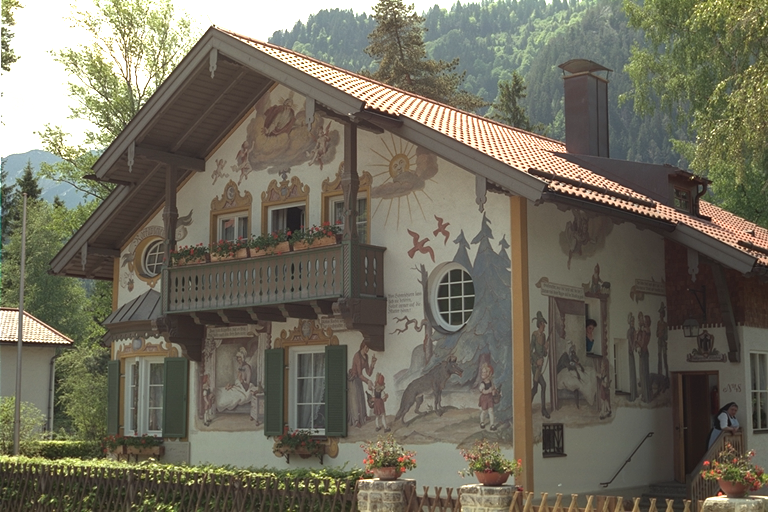}
\caption*{Ground Truth}
 \end{subfigure}
  \begin{subfigure}{0.33\linewidth}
\includegraphics[scale=0.21]{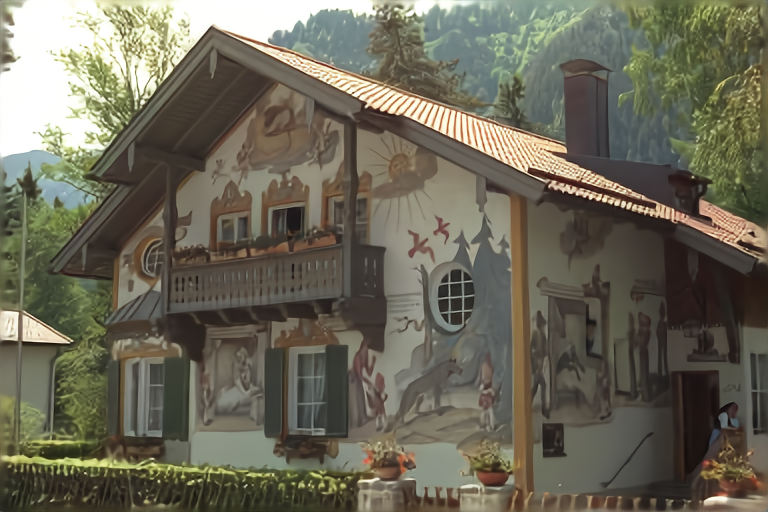}
\caption*{(a) 0.238bpp / 24.94dB / 0.9668}
 \end{subfigure}
  \begin{subfigure}{0.33\linewidth}
\includegraphics[scale=0.21]{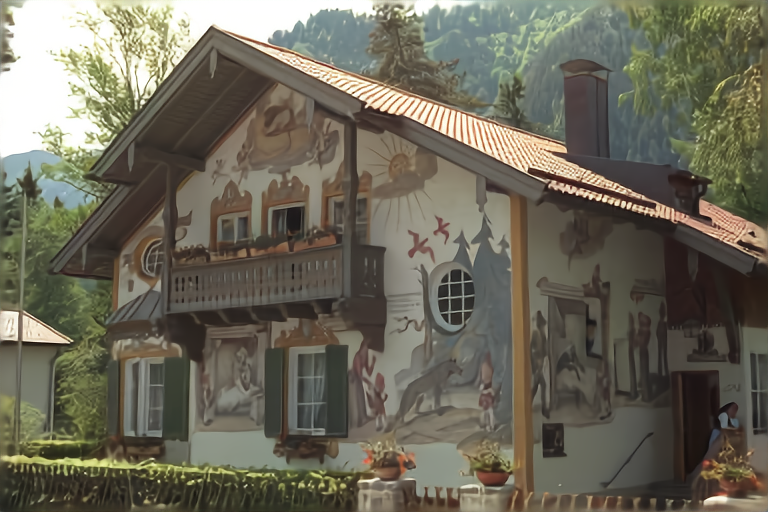}
\caption*{(b) 0.234bpp / 24.97dB / \textbf{0.9673}}
 \end{subfigure}
 \caption{Visual comparisons. (a) Base model \cite{cheng2020learned} optimized for MS-SSIM. (b) Our methods optimized for MS-SSIM.}
\label{figure12}
\end{figure*}

More reconstruction results are provided here for visual comparisons (Figure \ref{figure11} and Figure \ref{figure12}). The base model is still \cite{cheng2020learned}. And we show both the PSNR-optimized and the MS-SSIM-optimized results.

\end{document}